\documentstyle[12pt,preprint,aps,epsf,floats]{revtex}
\newcommand{\nn}{\nonumber}
\newcommand{\be}{\begin{equation}}
\newcommand{\ee}{\end{equation}}
\newcommand{\bea}{\begin{eqnarray}}
\newcommand{\eea}{\end{eqnarray}}

\def\bfnabla{\mbox{\boldmath $\nabla$}}

\def\bfsigma{\mbox{\boldmath $\sigma$}}
\def\bfPi{\mbox{\boldmath $\Pi$}}
\def\lQ{\Lambda_{\rm QCD}}
\def\al{\alpha}
\def\als{\alpha_{\rm s}}
\def\siml{{\ \lower-1.2pt\vbox{\hbox{\rlap{$<$}\lower6pt\vbox{\hbox{$\sim$}}}}\ }} 
\def\vac{\hbox{vac}}
\def\lla{\langle\!\langle}
\def\rra{\rangle\!\rangle}
\newcommand{\Appendix}[1]%
    {%
     \section{#1}%
      }

\begin{document}
\preprint{\tt CERN-TH/2000-197, HD-THEP-00-31 \vspace{1cm}}
\title{\bf The QCD potential at $O(1/m^2)$: \\ 
 Complete spin-dependent and spin-independent result\vspace{1cm}}
\author{Antonio Pineda\footnote{antonio.pineda@cern.ch}}
\address{Theory Division CERN, 1211 Geneva 23, Switzerland}
\author{Antonio Vairo\footnote{antonio.vairo@cern.ch}}
\address{Institut f\"ur Theoretische Physik, Universit\"at Heidelberg\\ 
Philosophenweg 16, D-69120 Heidelberg, Germany}
\maketitle

{\tighten
\begin{abstract}
\baselineskip=20 pt 
\noindent 
Within an effective field theory framework, we obtain an expression, 
with $O(1/m^2)$ accuracy, for the energies of the gluonic excitations
between heavy quarks, which holds beyond perturbation theory. 
For the singlet heavy quark--antiquark energy, in particular, we also obtain 
an expression in terms of Wilson loops. This provides, twenty years after 
the seminal work of Eichten and Feinberg, the first complete expression 
for the heavy quarkonium potential up to $O(1/m^2)$ for pure gluodynamics. 
Several errors present in the previous literature (also in the work of Eichten
and Feinberg) have been corrected. We also briefly discuss the power counting
of NRQCD in the non-perturbative regime.  
\end{abstract}

\vfill
\eject                                                              

\tighten

\section{Introduction}
The measured spectroscopy suggests that the charm and bottom quark masses are
large enough to consider their heavy-quark--antiquark bound-state systems
(generically denoted as heavy quarkonia: $\psi$, $\Upsilon$, $B_c$, ...) as
non-relativistic (NR).  These systems are, therefore, characterized by, at
least, three widely separated scales: hard (the mass $m$ of the heavy
quarks), soft (the relative momentum of the heavy-quark--antiquark $|{\bf p}|
\sim mv$, $ v \ll 1$), and ultrasoft (the typical kinetic energy $E \sim mv^2$
of the heavy quark in the bound-state system).  Inspired by this NR behaviour,
the investigation of heavy quarkonia has been traditionally performed by all
sorts of potential models, where an {\it ansatz} potential is introduced in a
Schr\"odinger equation (for some reviews see \cite{rev0,rev1,gunnar}).  The
phenomenological success of these suggests that, to some extent, a potential
picture may, in fact, be appropriate and justified from QCD. This triggered
the attempts to derive these potentials from QCD by relating them to Wilson
loops.  These standard derivations used an expansion in $1/m$ (also named
adiabatic or Born--Oppenheimer approximation).  However, a full derivation of
the potential from QCD, as well as a study of the validity of the potential
picture itself, was not done so far in the non-perturbative regime, where most
of the heavy quarkonium spectrum lies. It is the aim of this paper to
explicitly derive the complete non-perturbative $1/m^2$ QCD potential for pure
gluodynamics within an effective field theory framework \cite{pNRQCD,m1},
where higher order potentials in $1/m$ and non-potential effects could also be
incorporated in a systematic way.

Since the derivation of the potential has a long story, it may be useful to
summarize its main steps. The expression for the leading spin-independent
potential, of $O(1/m^0)$, corresponds to the static Wilson loop and was
derived and discussed in the seminal works of Wilson and Susskind
\cite{wilson,Brown}.  Expressions for the leading spin-dependent potentials in
the $1/m$ expansion, of $O(1/m^2)$, were given in Refs. \cite{spin1,Peskin,spin2}.
The procedure followed in these works proved to be very difficult to extend beyond these
leading-order potentials.  Indeed, the first attempts \cite{thesis}, using
tools similar to those in Ref. \cite{spin1}, failed to obtain suitable finite
expressions.  In Ref. \cite{BMP}, a new method to calculate the potentials
was proposed, where new spin-independent (some of them momentum-dependent)
potentials at $O(1/m^2)$ were obtained.  In these original works, the obtained
potentials did not correctly reproduce the ultraviolet behaviour expected from
perturbative QCD (the hard logs $\sim \log m$). This was first implemented in
the framework of QCD effective field theories, for both spin-dependent and
spin-independent potentials, in \cite{chen,latpot,BV1}.  At that point, the
obtained set of potentials at $O(1/m^2)$ seemed to be complete and 
the timely study of the different Wilson loop operators describing the
non-perturbative dynamics of the potentials started. For instance, a lattice
study was performed in \cite{latpot} and a study in the framework of QCD
vacuum models was done in \cite{mod}.

Nevertheless, this view has been recently challenged in Ref. \cite{m1} where:
{\it i)} a systematic study of the potentials has been started within an
effective field theory framework: potential-NRQCD (pNRQCD) \cite{pNRQCD}, and
{\it ii)} the $O(1/m)$ potential, previously missed in the literature, has
been calculated.  It is the aim of this paper to explain in more detail the
Hamiltonian formalism, sketched in Ref. \cite{m1}, and to compute the
$O(1/m^2)$ potentials. The formalism appears to be quite powerful and suitable
to obtain the quarkonium potentials and the energies of any gluonic excitation
at any finite order in $1/m$. A similar idea, but in the Coulomb gauge and
only for the leading spin-dependent quarkonium potentials, has also been used
in \cite{SS}. We will give an expression in terms of quantum-mechanical
corrections to the energies of the gluonic excitations between static quarks,
valid for all the gluonic excitations up to $O(1/m^2)$.  For the quarkonium
state (the ground state), we will express our complete $1/m^2$ result in terms
of Wilson loops eventually calculable on the lattice or by means of QCD vacuum models,
concluding in this way an ideal journey started over twenty years ago.

The theoretical framework of our work is NRQCD \cite{NRQCD} and pNRQCD,
suitable effective field theories for systems made up by two heavy quarks.
NRQCD has proved to be extremely successful in studying heavy quark--antiquark
systems near threshold. It is obtained from QCD by integrating out the hard
scale $m$.  It is characterized by an ultraviolet cut-off much smaller than
the mass $m$ and much larger than any other scale, in particular much larger
than $\lQ$.  This means that the matching from QCD to NRQCD can always be done
perturbatively, as well as within an expansion in $1/m$ \cite{Manohar,Match}.
The Lagrangian of NRQCD can also be organized in powers of $1/m$, thus making 
explicit the non-relativistic nature of the physical systems. So far, NRQCD
and pNRQCD have only been studied in detail in the perturbative situation
\cite{others,pNRQCD}.

By integrating out degrees of freedom with energies larger than $mv^2$, one is
left to a new effective field theory called pNRQCD where the soft and
ultrasoft scales have been disentangled and where the connection between NRQCD
and a NR quantum-mechanical description of the system can be formalized in a
systematic way. pNRQCD has two ultraviolet cut-offs, $\Lambda_1$ and
$\Lambda_2$.  The former fulfils the relation $ mv^2$ $\ll \Lambda_1 \ll$ $mv$
and is the cut-off of the energy of the quarks, and of the energy and the
momentum of the gluons, whereas the latter fulfils $mv \ll \Lambda_2 \ll m$
and is the cut-off of the relative momentum of the quark--antiquark system,
${\bf p}$. In the non-perturbative situation (we understand by
non-perturbative a typical situation where $mv \sim \lQ$, i.e.  where the
potential cannot be computed perturbatively), we will assume that the matching
between NRQCD and pNRQCD can be performed, as in the perturbative case, order
by order in the $1/m$ expansion.  We will present, for the general situation
$\lQ \siml mv$, the matching of NRQCD to pNRQCD at $O(1/m^2)$ for the singlet
sector (to be defined later). This will prove to be equivalent to computing the
heavy quarkonium potential that we can now derive from QCD by a systematic
procedure. Moreover, the expression for the potential that we obtain will also
be correct at any power in $\als$ in the perturbative regime.

A pure potential picture emerges in pure gluodynamics under the condition
that all gluonic excitations have a gap larger than $m v^2$.  Extra
ultrasoft degrees of freedom such as hybrids and pions can be
systematically included and may eventually affect the leading potential
picture (as ultrasoft gluons in the perturbative regime \cite{pNRQCD}). 

In this paper we consider the general situation of particles with different
masses. Therefore, our results, besides to the traditional $Q$-${\bar Q}$
systems, may be applied to the $B_c$ system, which, after its recent discovery
by the CDF collaboration \cite{bcexp}, has received a lot of attention in
theoretical investigations \cite{bc}.

The paper is organized in the following way. In section \ref{secnrqcd} we
introduce NRQCD up to $O(1/m^2)$. In section \ref{secqm}, using a Hamiltonian
formulation of NRQCD, we explicitly calculate up to $O(1/m^2)$ the energies of
the gluonic excitations between heavy quarks. In section \ref{secpnrqcd} we
define what pNRQCD will be in the present context.  In section \ref{secwilson}
we write the heavy quarkonium potential up to $O(1/m^2)$ in terms of Wilson
loops and compare with previous results. In section \ref{secpc} we discuss the
power counting of pNRQCD in the non-perturbative regime and in section
\ref{conclusions} we give our conclusions and outlook.

\section{NRQCD}
\label{secnrqcd}
After integrating out the hard scale $m$, one obtains NRQCD \cite{NRQCD}. 
Neglecting operators that involve light quark fields \cite{ManBauer},
the most general NRQCD Lagrangian (up to field redefinitions) for a quark of mass $m_1$ 
and an antiquark of mass $m_2$ up to $O(1/m^2)$ is given by:
\bea
\label{lNRQCD}
&&{\cal L}_{\rm NRQCD}= \psi^\dagger \Biggl\{ i D_0 + \, {{\bf D}^2\over 2 m_1}
+ c^{(1)}_F\, g {{\bf \bfsigma \cdot B} \over 2 m_1}
+ c^{(1)}_D \, g { \left[{\bf D} \cdot, {\bf E} \right] \over 8 m_1^2}
+ i c^{(1)}_S \, g { {\bf \bfsigma \cdot \left[D \times, E \right] }\over 8 m_1^2} \Biggr\} \psi \\ \nn 
&& \qquad\qquad 
+ \chi^\dagger \Biggl\{ i D_0 - \, {{\bf D}^2\over 2 m_2} 
- c^{(2)}_F\, g {{\bf \bfsigma \cdot B} \over 2 m_2}
+ c^{(2)}_D \, g { \left[{\bf D \cdot, E} \right] \over 8 m_2^2}
+ i c^{(2)}_S \, g { {\bf \bfsigma \cdot \left[D \times, E\right] }\over 8 m_2^2} \Biggr\} \chi \\ \nn
&&
+ {d_{ss} \over m_1m_2} \psi^{\dag} \psi \chi^{\dag} \chi
+ {d_{sv} \over m_1m_2} \psi^{\dag} {\bfsigma} \psi \chi^{\dag} {\bfsigma} \chi
+ {d_{vs} \over m_1m_2} \psi^{\dag} {\rm T}^a \psi \chi^{\dag} {\rm T}^a \chi
+ {d_{vv} \over m_1m_2} \psi^{\dag} {\rm T}^a {\bfsigma} \psi \chi^{\dag} {\rm T}^a {\bfsigma} \chi \\ \nn
&&
- {1\over 4} G^a_{\mu \nu} G^{a\,\mu \nu} 
+ \left({d_2^{(1)}\over m_1^2} + {d_2^{(2)}\over m_2^2}\right) G^a_{\mu \nu} D^2 G^{a\,\mu \nu}
+ \left({d_3^{(1)}\over m_1^2} + {d_3^{(2)}\over m_2^2}\right) 
g f_{abc}G^a_{\mu\nu} G^b_{\mu\al} G^c_{\nu\al}, 
\nn
\eea
where $\psi$ is the Pauli spinor field that annihilates the fermion and $\chi$
is the Pauli spinor field that creates the antifermion,
$i D^0=i\partial_0 -gA^0$, $i{\bf D}=i\bfnabla+g{\bf A}$,
$[{\bf D \cdot, E}]={\bf D \cdot E} - {\bf E \cdot D}$ and 
$[{\bf D \times, E}]={\bf D \times E -E \times D}$.
This Lagrangian is sufficient to obtain the $O(1/m^2)$ potentials. 
The coefficients $c_F$, $c_D$, $c_S$, $d_2$ and $d_3$ 
can be found in Ref. \cite{Manohar} and $d_{ij}$ ($i,j=s,v$) in
\cite{Match} for the $\overline{MS}$ scheme.

Some words of caution are in order here.  Even if the above matching
coefficients have been computed using dimensional regularization and the
$\overline{MS}$ scheme, there could still remain some ambiguity depending on
the different prescriptions for the $\epsilon^{ijk}$ tensors and the
definition of the Pauli matrices $\bfsigma$. For instance, the use of a scheme
where the $\epsilon^{ijk}$ only takes values for dimension equal to three ('t
Hooft--Veltmann-like scheme) in the computation of Ref. \cite{Match} would
change the value of $d_{vv}$ as $d_{vv} \rightarrow \displaystyle {2 \over
  D-2} d_{vv}$, where $D$ is the number of space-time dimensions. One should therefore
be careful and make sure that the matching coefficients one is working with
really are computed in the same scheme. A deep study of these
ambiguities in the framework of NRQCD remains to be done. This may be specially
important for higher order calculations. See also Refs. \cite{CMYm6,ai1},
where the authors have to deal with equivalent problems.

We are interested in the Hamiltonian of the above Lagrangian. The construction of the
Hamiltonian of one effective (non-renormalizable) Lagrangian may be
complicated (for a related discussion we refer to \cite{knetter});
in particular because there are higher time derivatives acting on the
different fields. In order to get rid of those at $O(1/m^2)$ we have to
eliminate the term $G^a_{\mu \nu} D^2 G^{a\,\mu \nu}$ from the Lagrangian.
This can be achieved by a field redefinition as follows.
We consider the field redefinition of the gluon field ($c \sim 1/m^2$): 
\be
A_{\mu} \rightarrow A_{\mu} + c [D^\al,G_{\al \mu}]+O(c^2) \label{fred} \,,
\ee
where $c$ is real. This transformation preserves the gauge transformation properties and the
hermiticity of the $A_{\mu}$ field. Eq. (\ref{fred}) produces the following change in the
gluon Lagrangian (at the order of interest):
\be
- {1\over 4} G^a_{\mu \nu} G^{a\,\mu \nu} \rightarrow - {1\over 4} G^a_{\mu \nu}
G^{a\,\mu \nu} -{c \over 2}G_{\mu \nu}^a D^2 G^{a\,\mu \nu} 
-c gf_{abc}G_{\mu \nu}^aG_{\mu \alpha}^bG_{\nu \alpha}^c
+O(c^2) \,.
\ee
We can therefore cancel the $GD^2G$ term by fixing
\be
c={2d_2^{(1)} \over m_1^2}+{2d_2^{(2)} \over m_2^2}.
\ee
This changes the value of $d_3$ to $d_3^\prime$:
\be
d_3^{(1)\prime} =d_3^{(1)}-2d_2^{(1)}\,, \quad \quad d_3^{(2)\prime}
=d_3^{(2)}-2d_2^{(2)}
\,.
\ee
Let us now see the modifications that the above field redefinition will produce in other sectors 
of the theory, in particular, in the heavy fermion bilinear Lagrangian. Since we have the following 
change for the $D_0$  covariant derivative that appears at $O(1/m^0)$
\be
iD_0 \rightarrow iD_0 - c g[{\bf D} \cdot, {\bf E}] \,,
\ee
the matching coefficients change, at $O(1/m^2)$, as (all the others remain unchanged):
\be
c_D^{(1)\prime}= c_D^{(1)} -16 d_2^{(1)}-16 {m_1^2 \over m_2^2}d_2^{(2)}
\,,
\quad
c_D^{(2)\prime}= c_D^{(2)} -16 d_2^{(2)}-16 {m_2^2 \over m_1^2}d_2^{(1)} \,.
\ee
In summary, eliminating the term $G^a_{\mu \nu} D^2 G^{a\,\mu \nu}$, up to order $1/m^2$, 
is equivalent to the redefinition of the matching coefficients $d_3 \to d_3^\prime$ and 
$c_D \to c_D^\prime$ found above. We will assume this field redefinition in the following.

\section{Gluonic Excitations in a Hamiltonian Formulation}
\label{secqm}
The Hamiltonian associated to the Lagrangian (\ref{lNRQCD}) is, up to order $1/m^2$,
\begin{eqnarray}
H &=& H^{(0)}+{1 \over m_1}H^{(1,0)}+{1 \over m_2}H^{(0,1)} 
+{1 \over m_1^2}H^{(2,0)} +{1 \over m_2^2}H^{(0,2)}+{1 \over m_1m_2}H^{(1,1)},
\label{HH}\\
H^{(0)} &=& \int d^3{\bf x} {1\over 2}\left( {\bfPi}^a{\bfPi}^a +{\bf B}^a{\bf B}^a \right),
\label{H0}\\
H^{(1,0)} &=& - {1 \over 2} \int d^3{\bf x} \psi^\dagger \left( {\bf D}^2 
+ g c_F^{(1)} \bfsigma \cdot {\bf B}\right) \psi,  
\qquad 
H^{(0,1)} = {1 \over 2} \int d^3{\bf x} \chi^\dagger \left({\bf D}^2 
+ g c_F^{(2)} \bfsigma \cdot {\bf B} \right) \chi,
\label{H1}\\
H^{(2,0)} &=& \int \! d^3{\bf x}\,
\psi^\dagger\Biggl\{ - c_D^{(1)\prime} \, g { \left[{\bf D} \cdot, {\bf E}  \right] \over 8 }
- i c_S^{(1)} \, g {  \bfsigma \cdot \left[{\bf D} \times, {\bf E} \right] \over 8 } \Biggr\} \psi 
\! - \! \int d^3{\bf x} \, d_3^{(1)\prime} g f_{abc}G^a_{\mu\nu} G^b_{\mu\al} G^c_{\nu\al},
\label{H20}\\ 
H^{(0,2)} &=& H^{(2,0)}(\psi \leftrightarrow \chi; 1 \leftrightarrow 2),
\label{H02}\\ 
H^{(1,1)} &=& - \! \int \!\! d^3{\bf x} \left( 
  d_{ss}  \psi^{\dag} \psi \chi^{\dag} \chi
+ \! d_{sv}  \psi^{\dag} {\bfsigma} \psi \chi^{\dag} {\bfsigma} \chi
+ \! d_{vs}  \psi^{\dag} {\rm T}^a \psi \chi^{\dag} {\rm T}^a \chi
+ \! d_{vv}  \psi^{\dag} {\rm T}^a {\bfsigma} \psi \chi^{\dag} {\rm T}^a {\bfsigma} \chi \right),  
\label{H11}
\end{eqnarray}
where $\bfPi^a$ is the canonical momentum conjugated to ${\bf A}^a$ 
and the physical states are constrained to satisfy the Gauss law:
\be
{\bf D}\cdot {\bfPi}^a \vert {\rm phys} \rangle = 
g (\psi^\dagger T^a \psi + \chi^\dagger T^a \chi) \vert {\rm phys} \rangle.
\label{gausslaw}
\ee
Since $\bfPi^a = {\bf E}^a + O(1/m^2)$, in Eqs. (\ref{H20}-\ref{H02}) and in the rest of the paper, we will 
use the chromoelectric field instead of the canonical momentum where, to the order 
we are interested in, it does not affect our results.

\subsection{The static limit}
We are interested in the one-quark--one-antiquark sector of the Fock space. 
In the static limit the one-quark--one-antiquark sector of the Fock space can be spanned by
\begin{equation}
\vert \underbar{n}; {\bf x}_1 ,{\bf x}_2  \rangle^{(0)}
\equiv  \psi^{\dagger}({\bf x}_1) \chi_c^{\dagger} ({\bf x}_2)
|n;{\bf x}_1 ,{\bf x}_2\rangle^{(0)},\qquad \forall {\bf x}_1,{\bf x}_2\,,
\label{basis0}
\end{equation}
where $|\underbar{n}; {\bf x}_1 ,{\bf x}_2\rangle^{(0)} $ is a gauge-invariant
eigenstate (up to a phase) of $ H^{(0)}$, as a consequence of the Gauss law, 
with energy $E_{n}^{(0)}({\bf x}_1 ,{\bf x}_2)$. For convenience, we use here the field
$\chi_c ({\bf x})=i\sigma^2 \chi^{*} ({\bf x})$, instead of
$\chi ({\bf x})$, because it is the one to which a particle interpretation
can be easily given: it corresponds to a Pauli spinor that annihilates a fermion in 
the $3^*$ representation of color $SU(3)$ with the standard, particle-like, spin structure.
$|n;{\bf x}_1 ,{\bf x}_2\rangle^{(0)}$ encodes the gluonic content of the
state, namely it is annihilated by $\chi_c({\bf x})$ and $\psi ({\bf x})$ ($\forall {\bf x}$). 
It transforms as a $3_{{\bf x}_1}\otimes 3_{{\bf x}_2}^{\ast}$ under colour $SU(3)$. 
The normalizations are taken as follows
$$
^{(0)}\langle m;{\bf x}_1 ,{\bf x}_2|n;{\bf x}_1 ,{\bf x}_2\rangle^{(0)}  =\delta_{nm},
$$
$$
^{(0)}\langle \underbar{m}; {\bf x}_1 ,
{\bf x}_2|\underbar{n}; {\bf y}_1 ,{\bf y}_2\rangle^{(0)} =\delta_{nm}
\delta^{(3)} ({\bf x}_1 -{\bf y}_1)\delta^{(3)} ({\bf x}_2 -{\bf y}_2)\,.
$$
We have made it explicit that the positions ${\bf x}_1$ and ${\bf x}_2$ of the quark and antiquark 
respectively are good quantum numbers for the static solution 
$|\underbar{n};{\bf x}_1 ,{\bf x}_2 \rangle^{(0)}$,  whereas $n$ generically denotes the remaining 
quantum numbers, which are classified by the irreducible 
representations of the symmetry group $D_{\infty h}$ (substituting the parity
generator by CP). We also choose the basis such that $T|\underbar{n};{\bf x}_1
,{\bf x}_2 \rangle^{(0)}= |\underbar{n};{\bf x}_1
,{\bf x}_2 \rangle^{(0)}$ where $T$ is the time-inversion operator. 
The ground-state energy $E_0^{(0)}({\bf x}_1,{\bf x}_2)$ can be 
associated to the static potential of the heavy quarkonium under some
circumstances (see Sec. \ref{secpnrqcd}). The remaining energies $E_n^{(0)}({\bf x}_1,{\bf x}_2)$, 
$n\not=0$, are usually associated to the potential used in order to describe 
heavy hybrids or heavy quarkonium (or other heavy hybrids) plus glueballs 
(see Sec. \ref{secpnrqcd}). They can be computed on the lattice 
(see for instance \cite{michael}). Translational invariance implies that  
$E_n^{(0)}({\bf x}_1,{\bf x}_2) = E_n^{(0)}(r)$, where ${\bf r}={\bf x}_1-{\bf x}_2$.

\subsection{Beyond the static limit}
Beyond the static limit, but still working order by order in $1/m$,
the normalized eigenstates, $|\underbar{n}; {\bf x}_1 ,{\bf x}_2\rangle$,
and eigenvalues, $E_n({\bf x}_1 ,{\bf x}_2; {\bf p}_1, {\bf p}_2)$, 
of the Hamiltonian $H$ satisfy the equations
\bea 
& & H |\underbar{n}; {\bf x}_1 ,{\bf x}_2\rangle = \int d^3x_1^\prime d^3x_2^\prime 
|\underbar{n}; {\bf x}_1^\prime ,{\bf x}_2^\prime \rangle 
E_n({\bf x}_1^\prime,{\bf x}_2^\prime; {\bf p}_1^\prime, {\bf p}_2^\prime)
\delta^{(3)}({\bf x}_1^\prime-{\bf x}_1)\delta^{(3)}({\bf x}_2^\prime-{\bf x}_2), \label{bornschroe}
\\
& &
\langle \underbar{m}; {\bf x}_1 ,{\bf x}_2|\underbar{n}; {\bf y}_1 ,{\bf y}_2\rangle = 
\delta_{nm} \delta^{(3)} ({\bf x}_1 -{\bf y}_1)\delta^{(3)} ({\bf x}_2 -{\bf y}_2).
\label{bornnorm}
\eea 
Note that the positions ${\bf x}_1$ and ${\bf x}_2$ of the static solution
still label the states even if the position operator does not commute with $H$
beyond the static limit. We are interested in the eigenvalues $E_n$, which
should be understood as operators (instead of numbers, even though we call
them energies). This will match the operator interpretation within a
quantum-mechanical formulation that we will give to them in pNRQCD in the next
section.  In particular, we will see that $E_0$ corresponds to the
quantum-mechanical Hamiltonian of the heavy quarkonium (in some specific
situation). The other energies, $E_n$ for $n > 0$, are related to the
quantum-mechanical Hamiltonians of the heavy hybrids or heavy quarkonium 
(or other heavy hybrids) plus glueballs.

Since the derivation of the corrections to $E_n$ may not be familiar to the
reader, since they are operators, we explain it in some
detail. We will work in the same way as in standard quantum mechanics, but
taking into account the fact that they are operators. Analogously to standard
quantum mechanics, we define a state $|\underbar{$\tilde n$}; {\bf x}_1 ,{\bf
  x}_2\rangle$ such that
\begin{eqnarray*}
& & H |\underbar{$\tilde n$}; {\bf x}_1 ,{\bf x}_2\rangle = \int d^3x_1^\prime d^3x_2^\prime 
|\underbar{$\tilde n$}; {\bf x}_1^\prime ,{\bf x}_2^\prime \rangle 
{\tilde E}_n({\bf x}_1^\prime,{\bf x}_2^\prime; {\bf p}_1^\prime, {\bf p}_2^\prime)
\delta^{(3)}({\bf x}_1^\prime-{\bf x}_1)\delta^{(3)}({\bf x}_2^\prime-{\bf x}_2), 
\\
& &
^{(0)}\langle \underbar{n}; {\bf x}_1 ,{\bf x}_2|\underbar{$\tilde n$}; {\bf y}_1 ,{\bf y}_2\rangle = 
\delta^{(3)} ({\bf x}_1 -{\bf y}_1)\delta^{(3)} ({\bf x}_2 -{\bf y}_2). 
\end{eqnarray*}
Splitting the Hamiltonian as $H=H_0+H_I$ we have 
\begin{eqnarray*}
& &\!\!\!\!\!\!\!\!
|\underbar{$\tilde n$}; {\bf x}_1 ,{\bf x}_2\rangle = 
|\underbar{n}; {\bf x}_1 ,{\bf x}_2\rangle^{(0)} + {1\over E_n^{(0)}(x)-H^{(0)}}
\sum_{m\not=n}\int d^3x_1^\prime d^3x_2^\prime
|\underbar{m}; {\bf x}_1^\prime ,{\bf x}_2^\prime\rangle^{(0)}
{}^{(0)}\langle \underbar{m}; {\bf x}_1^\prime ,{\bf x}_2^\prime|
\\
& &\times 
\bigg\{H_I |\underbar{$\tilde n$}; {\bf x}_1 ,{\bf x}_2\rangle  - \int d^3x_1^\prime d^3x_2^\prime
|\underbar{$\tilde n$}; {\bf x}_1^\prime ,{\bf x}_2^\prime \rangle 
\Delta{\tilde E}_n({\bf x}_1^\prime,{\bf x}_2^\prime; {\bf p}_1^\prime,
{\bf p}_2^\prime)
\delta^{(3)}({\bf x}_1^\prime-{\bf x}_1)\delta^{(3)}({\bf x}_2^\prime-{\bf x}_2) \bigg\},
\end{eqnarray*}
and
$$
\Delta {\tilde E}_n ({\bf x}_1,{\bf x}_2; {\bf p}_1, {\bf p}_2)
\delta^{(3)} ({\bf x}_1 -{\bf y}_1)\delta^{(3)}({\bf x}_2 -{\bf y}_2)
= {}^{(0)}\langle \underbar{n}; {\bf x}_1 ,{\bf x}_2| H_I 
|\underbar{$\tilde n$}; {\bf y}_1 ,{\bf y}_2\rangle. 
$$
From these formulas we can obtain ${\tilde E}_n$ order by order in the expansion parameter of $H_I$. 
Moreover $|\underbar{n}; {\bf x}_1 ,{\bf x}_2\rangle$ and $E_n$ are given by 
$$
|\underbar{n}; {\bf x}_1 ,{\bf x}_2\rangle =
\int d^3x_1^\prime d^3x_2^\prime |\underbar{$\tilde n$}; {\bf x}_1^\prime ,{\bf x}_2^\prime\rangle
N_n^{-1/2}({\bf x}_1^\prime,{\bf x}_2^\prime; {\bf p}_1^\prime, {\bf p}_2^\prime)
\delta^{(3)}({\bf x}_1^\prime-{\bf x}_1)\delta^{(3)}({\bf x}_2^\prime-{\bf x}_2),
$$
and
$$
E_n=N_n^{1/2}{\tilde E}_nN_n^{-1/2},
$$
where
$$
\langle \underbar{$\tilde n$}; {\bf x}_1 ,{\bf x}_2 |\underbar{$\tilde n$}; {\bf y}_1 ,{\bf y}_2\rangle
= N_n({\bf x}_1,{\bf x}_2; {\bf p}_1, {\bf p}_2)
\delta^{(3)} ({\bf x}_1 -{\bf y}_1)\delta^{(3)} ({\bf x}_2 -{\bf y}_2).
$$
By using the above results, we get for $E_n$ up to $O(1/m^2)$: 
\bea
& & E_n({\bf x}_1,{\bf x}_2; {\bf p}_1, {\bf p}_2)
\delta^{(3)}({\bf x}_1-{\bf x}_1^\prime)\delta^{(3)}({\bf x}_2-{\bf x}_2^\prime) =
E_n^{(0)}({\bf x}_1,{\bf x}_2) \delta^{(3)}({\bf x}_1-{\bf x}_1^\prime)
\delta^{(3)}({\bf x}_2-{\bf x}_2^\prime)
\nn\\
& & ~~ +\, ^{(0)}\langle \underbar{n}; {\bf x}_1 ,{\bf x}_2 \vert
{H^{(1,0)}\over m_1} + {H^{(0,1)}\over m_2} 
+{H^{(2,0)}\over m_1^2} + {H^{(0,2)}\over m_2^2} + {H^{(1,1)}\over m_1m_2} 
\vert \underbar{n}; {\bf x}_1^\prime ,{\bf x}_2^\prime \rangle^{(0)} 
\nn \\
& &
~~ - \, {1\over 2}\sum_{k\neq n} \int d^3y_1 \, d^3y_2 \,
^{(0)}\langle \underbar{n}; {\bf x}_1 ,{\bf x}_2 \vert 
{H^{(1,0)}\over m_1} + {H^{(0,1)}\over m_2} 
\vert \underbar{k}; {\bf y}_1 ,{\bf y}_2\rangle^{(0)}\,\nn\\
& &  \qquad \qquad \qquad \qquad \quad \times
^{(0)}\langle \underbar{k}; {\bf y}_1,{\bf y}_2 \vert 
{H^{(1,0)}\over m_1} + {H^{(0,1)}\over m_2} 
\vert \underbar{n}; {\bf x}_1^\prime,{\bf x}_2^\prime \rangle^{(0)}
\nn\\
& & \qquad \qquad \times 
\left( {1\over E_k^{(0)}({\bf y}_1,{\bf y}_2) - E_n^{(0)}({\bf x}_1^\prime,{\bf x}_2^\prime)} 
+ {1\over E_k^{(0)}({\bf y}_1,{\bf y}_2) - E_n^{(0)}({\bf x}_1,{\bf x}_2)} \right).
\label{En2}
\eea  

The expansion of $E_n$ in inverse powers of the mass can be organized up to $O(1/m^2)$ as follows:
\be
E_n = E_n^{(0)}+{E_n^{(1,0)} \over m_1}+{E_n^{(0,1)} \over m_2}+ {E_n^{(2,0)} \over m_1^2}
+ {E_n^{(0,2)}\over m_2^2}+{E_n^{(1,1)} \over m_1m_2}. \label{Enexp}
\ee
From Eq. (\ref{En2}) and Eqs. (\ref{H1})--(\ref{H11}), by using the identities (here and in the rest of
the paper, if not explicitly stated, the dependence on ${\bf x}_1$ and ${\bf x}_2$ is understood): 
\begin{eqnarray*}
&\hbox{a)}&  \quad 
{}^{\,\,(0)} \langle n | {\bf D}_{1} | n \rangle^{(0)} = \bfnabla_{1}, \qquad\qquad\qquad  
{}^{\,\,(0)} \langle n | {\bf D}_{c \,2} | n \rangle^{(0)} = \bfnabla_{2}, \\
&\hbox{b)}& \quad 
{}^{\,\,(0)} \langle n | {\bf D}_{1} | j \rangle^{(0)} = 
{ {}^{\,\,(0)} \langle n | g{\bf E}_1 | j \rangle^{(0)} \over
E_n^{(0)} - E_j^{(0)}}, \quad 
{}^{\,\,(0)} \langle n | {\bf D}_{c\,2} | j \rangle^{(0)} = 
-{ {}^{\,\,(0)} \langle n | g{\bf E}^{T}_2 | j \rangle^{(0)} \over
E_n^{(0)} - E_j^{(0)}} ~\forall \, n\neq j,\\
&\hbox{c)}& \quad 
{}^{\,\,(0)} \langle n | g{\bf E}_1 | n \rangle^{(0)} = 
-(\bfnabla_{1} E_n^{(0)}), \qquad~ 
{}^{\,\,(0)} \langle n | g{\bf E}^T_2 | n \rangle^{(0)} = 
(\bfnabla_{2} E_n^{(0)}), 
\end{eqnarray*}
where $F_j \equiv F({\bf x}_j)$, $\bfnabla_{j}=\bfnabla_{{\bf x}_j}$, 
${\bf D}_{c\,j}=\bfnabla_{j}+ig{\bf A}^T_j$,  
and the transpose refers  to the color matrices, we obtain at $O(1/m)$:
\be
E_n^{(1,0)} = {1\over 2} \sum_{k\neq n} \left\vert{ {}^{\,\,(0)} \langle k | 
g{\bf E}_1 | n \rangle^{(0)} \over E_n^{(0)} - E_k^{(0)}} \right\vert^2, 
\qquad 
E_n^{(0,1)} = {1\over 2} \sum_{k\neq n} \left\vert{ {}^{\,\,(0)} \langle k | 
g{\bf E}_2^{T} | n \rangle^{(0)} \over E_n^{(0)} - E_k^{(0)}} \right\vert^2. 
\label{Em1}
\ee
By using translational invariance one can see that $E_n^{(1,0)}$ and $E_n^{(0,1)}$ 
only depend on the relative distance $r$. Moreover, by using the symmetries of the static solutions, 
we can also see that $E_n^{(1,0)}=E_n^{(0,1)}$.  
The expressions (\ref{Em1}) were first derived in Ref. \cite{m1}.

At $O(1/m^2)$, we obtain
\bea 
& & \hspace{-14mm}E_n^{(2,0)} = 
-{c_D^{(1)\prime} \over 8}
{}^{(0)}\langle n|[{\bf D}_1\cdot, g{\bf E}_1]|n\rangle^{(0)}
+ {c_F^{(1)\,2}\over 4}\sum_{k\neq n} 
{^{(0)}\langle n| g {\bf B}_1|k\rangle^{(0)} 
\cdot   \,
^{(0)}\langle k| g {\bf B}_1|n\rangle^{(0)}   \over E_n^{(0)} - E_k^{(0)}}\nn\\
& & \hspace{-8mm}
+ \! {1\over 2}\sum_{k\neq n} \left[ 
\left\{p^i_1p^j_1,{^{(0)}\langle n| g{\bf E}^i_1|k\rangle^{(0)} \, ^{(0)}\langle k| g{\bf E}^j_1|n\rangle^{(0)} 
\over (E_n^{(0)}-E_k^{(0)})^3} \right\} \!
+ \! \left(\nabla_1^i\nabla_1^j {^{(0)}\langle n| g{\bf E}^i_1|k\rangle^{(0)} \,  
^{(0)}\langle k| g{\bf E}^j_1|n\rangle^{(0)} 
\over (E_n^{(0)}-E_k^{(0)})^3}\right) \right.\nn\\
& & \hspace{-8mm}
+ 2 \sum_{j,\ell \neq n} { ^{(0)}\langle n| g{\bf E}^i_1|j\rangle^{(0)} \, 
^{(0)}\langle j| g{\bf E}^i_1|k\rangle^{(0)}   
\,^{(0)}\langle k| g{\bf E}^j_1|\ell\rangle^{(0)}\,  ^{(0)}\langle \ell| g{\bf E}^j_1|n\rangle^{(0)}  
\over (E_n^{(0)}-E_k^{(0)})^3(E_n^{(0)}-E_j^{(0)})(E_n^{(0)}-E_\ell^{(0)})} \nn\\
& & \hspace{-8mm}
+ 2 \left(\nabla_1^i\sum_{j\neq n} 
{ ^{(0)}\langle n| g{\bf E}^i_1|k \rangle^{(0)} \,   
^{(0)}\langle k| g{\bf E}_1|j \rangle^{(0)}  \cdot ^{(0)}\langle j | g{\bf E}_1|n\rangle^{(0)}  
\over (E_n^{(0)}-E_k^{(0)})^3(E_n^{(0)}-E_j^{(0)})} \right)\nn\\
& & \hspace{-8mm}
- \left(\nabla_1^i {^{(0)}\langle n| g{\bf E}^i_1|k\rangle^{(0)} \,  
^{(0)}\langle k| [{\bf D}_1\cdot, g{\bf E}_1]|n\rangle^{(0)}  
\over (E_n^{(0)}-E_k^{(0)})^3}\right) \nn\\
& & \hspace{-8mm}
+ 3 \left(\nabla_1^i 
{^{(0)}\langle n| g{\bf E}^i_1|k\rangle^{(0)}  \, ^{(0)}\langle k| g{\bf E}_1|n\rangle^{(0)}  
\cdot (\bfnabla_1E_n^{(0)}) \over (E_n^{(0)}-E_k^{(0)})^4} \right) \nn\\
& & \hspace{-8mm}
-2 \sum_{j\neq n}
{ ^{(0)}\langle n| g{\bf E}_1|j \rangle^{(0)}  \cdot   
^{(0)}\langle j| g{\bf E}_1|k \rangle^{(0)} \,
^{(0)}\langle k | [{\bf D}_1\cdot, g{\bf E}_1] |n\rangle^{(0)}
\over (E_n^{(0)}-E_k^{(0)})^3(E_n^{(0)}-E_j^{(0)})} \nn\\
& & \hspace{-8mm}
+ 6 \sum_{j\neq n}
{ ^{(0)}\langle n| g{\bf E}_1|j \rangle^{(0)}  \cdot   
^{(0)}\langle j| g{\bf E}_1|k \rangle^{(0)} \, ^{(0)}\langle k | g{\bf E}_1 |n\rangle^{(0)}  
\cdot (\bfnabla_1E_n^{(0)})  \over (E_n^{(0)}-E_k^{(0)})^4(E_n^{(0)}-E_j^{(0)})} \nn\\
& & \hspace{-8mm}
- 3 {^{(0)}\langle n| [{\bf D}_1\cdot, g{\bf E}_1] |k \rangle^{(0)}  \,   
^{(0)}\langle k | g{\bf E}_1 |n\rangle^{(0)}  
\cdot (\bfnabla_1E_n^{(0)})  \over (E_n^{(0)}-E_k^{(0)})^4} \nn\\
& & \hspace{-8mm}
+ 4 { (\bfnabla_1E_n^{(0)}) \cdot ^{(0)}\langle n|g{\bf E}_1 |k \rangle^{(0)} \,  
^{(0)}\langle k | g{\bf E}_1 |n\rangle^{(0)}  \cdot (\bfnabla_1E_n^{(0)}) 
\over (E_n^{(0)}-E_k^{(0)})^5} \nn\\
& & \hspace{-8mm}
\left. +{1\over 2}
{^{(0)}\langle n| [{\bf D}_1\cdot, g{\bf E}_1] |k \rangle^{(0)} \, 
^{(0)}\langle k | [{\bf D}_1\cdot, g{\bf E}_1] |n\rangle^{(0)}  
\over (E_n^{(0)}-E_k^{(0)})^3} \right] \nn\\
& & \hspace{-8mm}
- d_3^{(1)\prime} f_{abc} \int d^3{\bf x} \,  
g \, ^{(0)}\langle n | G^a_{\mu\nu}({\bf x}) G^b_{\mu\al}({\bf x}) 
G^c_{\nu\al}({\bf x}) |n \rangle^{(0)}  \nn \\
& & \hspace{-8mm}
+ {c_F^{(1)}\over 2}\sum_{k\neq n} 
\left\{ \nabla^i_1, {^{(0)}\langle n| g {\bf E}_1^i|k\rangle^{(0)}  \,
^{(0)}\langle k| \bfsigma_1\cdot g {\bf B}_1|n\rangle^{(0)} \over (E_n^{(0)} - E_k^{(0)})^2} \right\}
-i{c_S^{(1)}\over 4}{ 1 \over r}{d\,E_n^{(0)} \over dr} \bfsigma_1 \cdot ({\bf r} \times \bfnabla_1) ,
\label{En20} \\
E_n^{(0,2)} \!\! &=& \! E_n^{(2,0)}(g{\bf E}_1 \rightarrow 
-g {\bf E}_2^T,g{\bf B}_1 \rightarrow -g {\bf B}_2^T,\bfsigma_1 \rightarrow
\bfsigma_2, \bfnabla_1 \rightarrow \bfnabla_2, {\bf D}_1 
\rightarrow {\bf D}_{c\,2}, m_1 \leftrightarrow m_2),
\label{En02} 
\eea
and 
\bea 
& & \hspace{-15mm}
E_n^{(1,1)} = \sum_{k\neq n} \left[ 
-\left\{p^i_1p^j_2,{^{(0)}\langle n| g{\bf E}^i_1|k\rangle^{(0)}\,  ^{(0)}\langle k| g{\bf E}^{j\,T}_2|n\rangle^{(0)} 
\over (E_n^{(0)}-E_k^{(0)})^3} \right\} \right. \nn\\
& & \hspace{-8mm}
- \left(\nabla_1^i\nabla_2^j {^{(0)}\langle n| g{\bf E}^i_1|k\rangle^{(0)}\,  
^{(0)}\langle k| g{\bf E}^{j\,T}_2|n\rangle^{(0)} \over (E_n^{(0)}-E_k^{(0)})^3}\right) \nn\\
& & \hspace{-8mm}
+2 \sum_{j,\ell \neq n} { ^{(0)}\langle n| g{\bf E}^i_1|j\rangle^{(0)}\,  
^{(0)}\langle j| g{\bf E}^i_1|k\rangle^{(0)} \,  
^{(0)}\langle k| g{\bf E}^{j\,T}_2|\ell\rangle^{(0)} \, ^{(0)}\langle \ell| g{\bf E}^{j\,T}_2|n\rangle^{(0)}  
\over (E_n^{(0)}-E_k^{(0)})^3(E_n^{(0)}-E_j^{(0)})(E_n^{(0)}-E_\ell^{(0)})} \nn\\
& & \hspace{-8mm}
+ \left(\nabla_1^i\sum_{j\neq n} 
{ ^{(0)}\langle n| g{\bf E}^i_1|k \rangle^{(0)}   \, 
^{(0)}\langle k| g{\bf E}^T_2|j \rangle^{(0)}  \cdot ^{(0)}\langle j | g{\bf E}^T_2|n\rangle^{(0)}  
\over (E_n^{(0)}-E_k^{(0)})^3(E_n^{(0)}-E_j^{(0)})} \right)\nn\\
& & \hspace{-8mm}
- \left(\nabla_2^i\sum_{j\neq n} 
{ ^{(0)}\langle n| g{\bf E}_1|j \rangle^{(0)}  \cdot   
^{(0)}\langle j| g{\bf E}_1|k \rangle^{(0)} \, ^{(0)}\langle k | g{\bf E}^{i\,T}_2|n\rangle^{(0)}  
\over (E_n^{(0)}-E_k^{(0)})^3(E_n^{(0)}-E_j^{(0)})} \right)\nn\\
& & \hspace{-8mm}
+ {1\over 2}\left(\nabla_1^i
{^{(0)}\langle n| g{\bf E}^i_1|k\rangle^{(0)} \,
^{(0)}\langle k| [{\bf D}_{c\,2}\cdot, g{\bf E}^{T}_2]|n\rangle^{(0)}
\over (E_n^{(0)}-E_k^{(0)})^3}\right) \nn\\
& & \hspace{-8mm}
+  {1\over 2} \left(\nabla_2^i
{^{(0)}\langle 0| [{\bf D}_1\cdot, g{\bf E}_1]|k\rangle^{(0)} \,  
^{(0)}\langle k| g{\bf E}^{i\,T}_2|n\rangle^{(0)}  
\over (E_n^{(0)}-E_k^{(0)})^3}\right)\nn\\
& & \hspace{-8mm}
- {3\over 2} \left(\nabla_1^i 
{^{(0)}\langle n| g{\bf E}^i_1|k\rangle^{(0)} \,  ^{(0)}\langle k| g{\bf E}^T_2|n\rangle^{(0)}  
\cdot (\bfnabla_2E_n^{(0)}) \over (E_n^{(0)}-E_k^{(0)})^4} \right) \nn\\
& & \hspace{-8mm}
- {3\over 2}\left(\nabla_2^i  
{(\bfnabla_1E_n^{(0)})\cdot ^{(0)}\langle n| g{\bf E}_1|k\rangle^{(0)}   \,
^{(0)}\langle k| g{\bf E}^{i\,T}_2|n\rangle^{(0)}   
\over (E_n^{(0)}-E_k^{(0)})^4} \right)\nn\\
& & \hspace{-8mm}
+ \sum_{j\neq n}
{ ^{(0)}\langle n| g{\bf E}_1|j \rangle^{(0)}  \cdot   
^{(0)}\langle j| g{\bf E}_1|k \rangle^{(0)} \,  
^{(0)}\langle k | [{\bf D}_{c\,2}\cdot, g{\bf E}^{T}_2] |n\rangle^{(0)}  
\over (E_n^{(0)}-E_k^{(0)})^3(E_n^{(0)}-E_j^{(0)})} \nn\\
& & \hspace{-8mm}
- \sum_{j\neq n}
{ ^{(0)}\langle n|[{\bf D}_1\cdot, g{\bf E}_1] |k \rangle^{(0)}  \,   
^{(0)}\langle k| g{\bf E}^T_2|j \rangle^{(0)}  \cdot ^{(0)}\langle j | g{\bf E}^T_2|n\rangle^{(0)}  
\over (E_n^{(0)}-E_k^{(0)})^3(E_n^{(0)}-E_j^{(0)})} \nn\\
& & \hspace{-8mm}
- 3 \sum_{j\neq n}
{ ^{(0)}\langle n| g{\bf E}_1|j \rangle^{(0)}  \cdot   
^{(0)}\langle j| g{\bf E}_1|k \rangle^{(0)} \,  ^{(0)}\langle k | g{\bf E}^{T}_2 |n\rangle^{(0)}  
\cdot (\bfnabla_2E_n^{(0)})  \over (E_n^{(0)}-E_k^{(0)})^4(E_n^{(0)}-E_j^{(0)})} \nn\\
& & \hspace{-8mm}
+ 3 \sum_{j\neq n}
{ (\bfnabla_1E_n^{(0)}) \cdot ^{(0)}\langle n|g{\bf E}_1 |k \rangle^{(0)}  \,   
^{(0)}\langle k| g{\bf E}^T_2|j \rangle^{(0)}  \cdot ^{(0)}\langle j | g{\bf E}^T_2|n\rangle^{(0)}  
\over (E_n^{(0)}-E_k^{(0)})^4(E_n^{(0)}-E_j^{(0)})} \nn\\
& & \hspace{-8mm}
+{3\over 2}
{^{(0)}\langle n| [{\bf D}_1\cdot, g{\bf E}_1] |k \rangle^{(0)}  \, 
^{(0)}\langle k | g{\bf E}^{T}_2 |n\rangle^{(0)}  \cdot (\bfnabla_2E_n^{(0)})  
\over (E_n^{(0)}-E_k^{(0)})^4} \nn\\
& & \hspace{-8mm}
+{3\over 2}
{(\bfnabla_1E_n^{(0)}) \cdot ^{(0)}\langle n|g{\bf E}_1 |k \rangle^{(0)}    \,
^{(0)}\langle j | [{\bf D}_{c\,2}\cdot, g{\bf E}^{T}_2]|n\rangle^{(0)}  
\over (E_n^{(0)}-E_k^{(0)})^4} \nn\\
& & \hspace{-8mm}
- 4 { (\bfnabla_1E_n^{(0)}) \cdot ^{(0)}\langle n|g{\bf E}_1 |k \rangle^{(0)} \,  
^{(0)}\langle k | g{\bf E}^{T}_2 |n\rangle^{(0)}  \cdot (\bfnabla_2E_n^{(0)}) 
\over (E_n^{(0)}-E_k^{(0)})^5} \nn\\
& & \hspace{-8mm}
\left. -{1\over 2}
{^{(0)}\langle n| [{\bf D}_1\cdot, g{\bf E}_1] |k \rangle^{(0)} \, 
^{(0)}\langle k | [{\bf D}_{c\,2}\cdot, g{\bf E}^{T}_2]|n\rangle^{(0)}  
\over (E_n^{(0)}-E_k^{(0)})^3} \right] \nn\\
& & \hspace{-8mm}
+ (d_{ss} + d_{vs}{}^{(0)}\langle n| T_1^a T_2^{a\,T} |n \rangle^{(0)} ) 
\,\delta^{(3)}({\bf x}_1-{\bf x}_2) \nn\\
& & \hspace{-8mm}
- {c_F^{(1)}\over 2}\sum_{k\neq n} 
\left\{ \nabla^i_2, {^{(0)}\langle n| g {\bf E}_2^{i\,T}|k\rangle^{(0)}  \,
^{(0)}\langle k| \bfsigma_1\cdot g {\bf B}_1|n\rangle^{(0)} \over (E_n^{(0)} - E_k^{(0)})^2} \right\} \nn\\
& & \hspace{-8mm}
- {c_F^{(2)}\over 2}\sum_{k\neq n} 
\left\{ \nabla^i_1, {^{(0)}\langle n| g {\bf E}_1^i|k\rangle^{(0)}  \,
^{(0)}\langle k| \bfsigma_2\cdot g {\bf B}_2^T|n\rangle^{(0)} \over (E_n^{(0)} - E_k^{(0)})^2} \right\} \nn\\
& & \hspace{-8mm}
-{c_F^{(1)}c_F^{(2)}\over 2} \sum_{k\neq n} 
{^{(0)}\langle n| \bfsigma_1 \cdot g {\bf B}_1|k\rangle^{(0)}\,
^{(0)}\langle k| \bfsigma_2 \cdot g {\bf B}_2^T|n\rangle^{(0)} \over E_n^{(0)} - E_k^{(0)}} \nn\\
& & \hspace{-8mm}
- (d_{sv} \bfsigma_1\cdot\bfsigma_2 
+ d_{vv} {}^{(0)}\langle n| T_1^a\bfsigma_1\cdot T_2^{a\,T} \bfsigma_2 |n \rangle^{(0)} ) \,
\delta^{(3)}({\bf x}_1-{\bf x}_2). 
\label{En11}
\eea 
The above equations (\ref{Em1})--(\ref{En11}) give the energies of the gluonic excitations 
between heavy quarks within an expansion in $1/m$ up to $O(1/m^2)$. 
From these expressions, in the case of the ground state ($n=0$), we will
derive, in section \ref{secwilson}, the equivalent Wilson loop expressions.

A similar approach has been used in Ref. \cite{SS} in order to derive, from
the QCD Hamiltonian in the Coulomb gauge, the spin-dependent part of the
potential up to $O(1/m^2)$. However, the behaviour at scales of $O(m)$ was not
correctly incorporated there. If we take our NRQCD matching coefficients at
tree level and neglect the tree-level annihilation contributions in the
equal-mass case, we find agreement for the spin-dependent potentials (up to
some transpose color matrices). Nevertheless, our general expression
(\ref{En2}) differs from the one used in \cite{SS}, which, in general, will
not give the correct spin-independent potentials.  This has to do, in our
opinion, with the fact that in order to derive Eq. (\ref{En2}) one has to deal
with operators rather than with numbers.

\section{pNRQCD}
\label{secpnrqcd}
In the previous section we have studied the static limit of NRQCD and its
corrections within a $1/m$ expansion. Let us now connect those results with
pNRQCD.

In the static limit, the gap between different states at fixed ${\bf r}$ will
depend on the dimensionless parameter $\lQ r$. In a general situation, there
will be a set of states $\{n_{\rm us}\}$ such that $E_{n_{\rm us}}^{(0)}(r)
\sim mv^2$ for the typical $r$ of the actual physical system. We denote these
states as ultrasoft. The aim of pNRQCD is to describe the behaviour of the
ultrasoft states. Therefore, all the physical degrees of freedom with energies
larger than $mv^2$ will be integrated out from NRQCD in order to obtain
pNRQCD. It is in this context that one may work order by order in $1/m$ (in
particular for the kinetic energy), and the calculation of the previous
section becomes the matching calculation between NRQCD and pNRQCD and provides
a rigorous connection with the adiabatic approximation (this approximation is
implicit in all the attempts at deriving the non-perturbative potentials from
QCD we are aware of). Whereas this can be justified within a perturbative
framework, in the non-perturbative case, we cannot, in general, guarantee the
validity of the $1/m$ expansion and one may think of examples where certain
degrees of freedom cannot be integrated out in the $1/m$ expansion (see
\cite{DS}).  We believe that this possibility, which, to our knowledge, has
never been mentioned before, except in Ref. \cite{m1}, deserves further study.
Note that this does not have to do with the consideration of ultrasoft
effects, which, unlike in earlier approaches, can be readily incorporated
within our formalism.
 
In the perturbative situation $\lQ r \ll 1$, which has been studied in detail
in \cite{pNRQCD}, $\{n_{\rm us}\}$ corresponds to a heavy-quark--antiquark state, 
in either a singlet or an octet configuration, plus gluons and light fermions, 
all of them with energies of $O(mv^2)$.  In a non-perturbative situation, which we will 
generically denote by $\lQ r \sim 1$, it is not so clear what $\{n_{\rm us}\}$ is. 
One can think of different possibilities. Each of them will give, in principle, different 
predictions and, therefore, it should be possible to experimentally discriminate 
among them. In particular, one could consider the situation where, because of a mass gap 
in QCD, the energy splitting between the ground state and the first gluonic excitation 
is larger than $mv^2$, and, because of chiral symmetry breaking of QCD, Goldstone bosons 
(pions/kaons) appear. Hence, in this situation, $\{n_{\rm us}\}$ would be the ultrasoft 
excitations about the static ground state (i.e. the solutions of the corresponding Schr\"odinger 
equation), which will be named the singlet, plus the Goldstone bosons. If one switches off 
the light fermions (pure gluodynamics), only the singlet survives and pNRQCD reduces to 
a pure two-particle NR quantum-mechanical system, usually referred as a pure
potential model.

In this paper, we will study the pure singlet sector, with no reference to further ultrasoft 
degrees of freedom. In this situation, pNRQCD only describes the ultrasoft excitations about the
static ground state of NRQCD. In terms of static NRQCD eigenstates, this means that only  
$|\underbar{0}; {\bf x}_1 ,{\bf x}_2\rangle^{(0)}$ is kept as an explicit degree of freedom 
whereas $|\underbar{n}; {\bf x}_1 ,{\bf x}_2\rangle^{(0)}$ with  $n\not=0$ are integrated 
out\footnote{In fact, we are only integrating out states with energies 
larger than $mv^2$ and all the states with $n\not=0$ will be understood in
this way throughout the paper. Since, in practice, we are integrating over all the states, 
if we are in the situation where some states, different from the singlet, are ultrasoft, 
these have to be subtracted later on. This is analogous to what happens in the
perturbative situation, where the subtraction is done order by order in the
multipole expansion. In this situation our calculation should be understood as
the leading term in the multipole expansion.}.
This provides the only dynamical degree of freedom of the theory. It is described by means 
of a bilinear colour singlet field, $S({\bf x}_1,{\bf x}_2,t)$, which has the same quantum numbers 
and transformation properties under symmetries as the static ground state of NRQCD in the 
one-quark--one-antiquark sector. In the above situation, the Lagrangian of pNRQCD reads
\be
{\cal L}_{\rm pNRQCD} = S^\dagger 
\bigg( i\partial_0 -h_s({\bf x}_1,{\bf x}_2, {\bf p}_1, {\bf p}_2)\bigg) S, 
\label{pnrqcdl}
\ee
where $h_s$ is the Hamiltonian of the singlet (actually $h_s$ is only a
function of {\bf r}, ${\bf p}_1$, ${\bf p}_2$, which is analytic in the two
last operators but typically contains non-analyticities in {\bf r}), 
${\bf p}_1= -i \bfnabla_{{\bf x}_1}$ and ${\bf p}_2= -i \bfnabla_{{\bf x}_2}$. 
It has the following expansion up to order $1/m^2$: 
\be
h_s ({\bf x}_1,{\bf x}_2, {\bf p}_1, {\bf p}_2) = 
{{\bf p}^2_1\over 2 m_1} +{{\bf p}^2_2\over 2 m_2} 
+ V^{(0)}+{V^{(1,0)} \over m_1}+{V^{(0,1)} \over m_2}+ {V^{(2,0)} \over m_1^2}
+ {V^{(0,2)}\over m_2^2}+{V^{(1,1)} \over m_1m_2}.
\label{hss}
\ee

The integration of higher excitations is trivial using the basis 
$|\underbar{n}; {\bf x}_1 ,{\bf x}_2\rangle$ since, in this case, they are decoupled 
from $|\underbar{0}; {\bf x}_1 ,{\bf x}_2 \rangle$. 
Then, the matching of NRQCD to pNRQCD consists in renaming things in a way such
that pNRQCD reproduces the matrix elements of NRQCD for the ground state, and, in
particular, the energy. This fixes the matching condition 
\be
E_0({\bf x}_1,{\bf x}_2, {\bf p}_1, {\bf p}_2) = h_s({\bf x}_1,{\bf x}_2, {\bf p}_1, {\bf p}_2) .
\label{singlet}
\ee

Although our main concern in this paper is to provide a well-controlled
derivation of the potential for the heavy quarkonium, we would like to say a
few words about the expressions $E_n$ ($n\not=0$) we have found in the
previous section. In the static limit, the different $E_n^{(0)}$ ($n\not=0$)
are identified with the static potentials to be used in a Schr\"odinger equation 
to obtain the spectra of the bound systems composed of a heavy quark and an antiquark 
(plus glueballs) different from the heavy quarkonium such as, for instance, heavy hybrids.
This assignment is argued within the adiabatic approximation and corresponds
to what is actually done in lattice simulations \cite{michael}. In this
respect, since we have given a systematic method to obtain the corrections
to the energy within a $1/m$ expansion, the energies $E_n$ correspond to the
quantum-mechanical Hamiltonians of the different bound systems made by a heavy quark and 
an antiquark (up to glueballs) and the $1/m$ and $1/m^2$ terms should be understood as the
relativistic corrections to the static potentials. It is still an open problem
if this procedure is the sensible thing to do for heavy hybrids, if (and whichever)
other possibilities may occur, and if these potentials, like the heavy
quarkonium potential, may eventually be written in terms of Wilson loops. We
will not deal with these problems here, which, however, deserve further
investigations. We refer to \cite{gunnar} for related discussions.

\section{Heavy Quarkonium Potential and Wilson Loops}
\label{secwilson}
In this section we express the heavy-quarkonium potential in terms of
Wilson-loop operators. These kinds of expressions are quite convenient for
lattice simulations or for QCD-vacuum-model studies (see, for instance,
\cite{latpot,mod}).  We shall use the following definitions. The angular
brackets $\langle \dots \rangle$ will stand for the average value over the
Yang--Mills action, $W_\Box$ for the rectangular static Wilson loop of
dimensions $r\times T_W$:
$$
W_\Box \equiv {\rm P} \exp\left\{{\displaystyle - i g \oint_{r\times T_W} \!\!dz^\mu A_{\mu}(z)}\right\},
$$
and $\langle\!\langle \dots \rangle\!\rangle
\equiv \langle \dots W_\Box\rangle / \langle  W_\Box\rangle$; P is the path-ordering operator.
Moreover, we define the {\it connected} Wilson loop 
with $O_1(t_1)$, $O_2(t_2)$, ..., $O_n(t_n)$ operator insertions for 
$T_W/2 \ge t_1 \ge t_2 \ge \dots \ge t_n \ge -T_W/2$ by: 
\bea
&&\lla O_1(t_1)O_2(t_2)\rra_c= \lla O_1(t_1)O_2(t_2)\rra 
-\lla O_1(t_1)\rra\lla O_{2}(t_2)\rra ,
\\ 
&&\lla O_1(t_1)O_2(t_2)O_3(t_3)\rra_c= \lla O_1(t_1)O_2(t_2)O_3(t_3)\rra
\nn\\
& & -\lla O_1(t_1)\rra \lla O_{2}(t_{2})O_3(t_3)\rra_c
-\lla O_1(t_1)O_2(t_2)\rra_c\lla O_{3}(t_{3})\rra
-\lla O_1(t_1)\rra \lla O_{2}(t_{2})\rra \lla O_3(t_3)\rra ,
\\
&&\lla O_1(t_1)O_2(t_2)O_3(t_3)O_4(t_4)\rra_c= 
\lla O_1(t_1)O_2(t_2)O_3(t_3)O_4(t_4)\rra 
\nn\\
\nn
& &
-\lla O_1(t_1)\rra \lla O_2(t_2)O_3(t_3)O_4(t_4)\rra_c
-\lla O_1(t_1)O_2(t_2)\rra_c\lla O_3(t_3)O_4(t_4)\rra_c
\\
\nn
& & 
-\lla O_1(t_1)O_2(t_2) O_3(t_3)\rra_c\lla O_4(t_4)\rra
-\lla O_1(t_1)\rra \lla O_2(t_2)\rra \lla O_3(t_3)O_4(t_4)\rra_c
\\
\nn
& & 
-\lla O_1(t_1)\rra \lla O_2(t_2)O_3(t_3)\rra_c \lla O_4(t_4)\rra
-\lla O_1(t_1)O_2(t_2)\rra_c \lla O_3(t_3)\rra \lla O_4(t_4)\rra
\\
& & 
-\lla O_1(t_1)\rra \lla O_2(t_2)\rra\lla O_3(t_3)\rra \lla O_4(t_4)\rra,
\\
\nn
&&\qquad\qquad\qquad\qquad\qquad \cdots
\eea 
We also define in a short-hand notation 
\be
\lim_{T\rightarrow \infty} \equiv \lim_{T\rightarrow \infty}\lim_{T_W\rightarrow \infty},
\ee
where $T_W$ is the time length of the Wilson loop and $T$ the time length appearing in the
time integrals. By performing first the $T_W\rightarrow \infty$, the averages $\lla \dots \rra$ 
become independent of $T_W$ and thus invariant under global time translations. 

By using the matching condition (\ref{singlet}) and the quantum-mechanical expressions (\ref{Em1}), it has already 
been proved in \cite{m1} that the quarkonium singlet static potential and the $O(1/m)$ potential 
can be expressed in terms of Wilson loops with field strength insertions in it as 
\be 
V^{(0)}(r) = \lim_{T\to\infty}{i\over T} \ln \langle W_\Box \rangle,  
\label{v0}
\ee
\be
V^{(1,0)}(r)=
-{1 \over 2} \lim_{T\rightarrow \infty}\int_0^{T}dt \, t \, \lla g{\bf E}_1(t)\cdot g{\bf E}_1(0) \rra_c.
\label{Em12}
\ee
Owing to invariance under charge conjugation plus $m_1 \leftrightarrow m_2$ transformation we have 
$$
V^{(1,0)}(r) = V^{(0,1)}(r).
$$
The way to prove the equivalence of Eq. (\ref{Em12}) and Eq. (\ref{Em1}) has been discussed 
in Ref. \cite{m1}, where more details can be found. 
Here we only mention that this equivalence proof as well as the following ones can be done 
straightforwardly by inserting complete sets of intermediate states in the Wilson
loop operators and by explicitly computing the time integrals. 

Let us now consider the terms of $O(1/m^2)$.
It is convenient to split them in a spin-dependent and a spin-independent part. 
For the $V^{(2,0)}$ and $V^{(0,2)}$ potentials we define
\be
V^{(2,0)}=V^{(2,0)}_{SD}+V^{(2,0)}_{SI}\,, \qquad V^{(0,2)}=V^{(0,2)}_{SD}+V^{(0,2)}_{SI}.
\ee
The spin-independent terms can be written as 
\be
V^{(2,0)}_{SI}={1 \over 2}\left\{{\bf p}_1^2,V_{{\bf p}^2}^{(2,0)}(r)\right\}
+{V_{{\bf L}^2}^{(2,0)}(r)\over r^2}{\bf L}_1^2 + V_r^{(2,0)}(r),
\ee
and
\be
V^{(0,2)}_{SI}={1 \over 2}\left\{{\bf p}_2^2,V_{{\bf p}^2}^{(0,2)}(r)\right\}
+{V_{{\bf L}^2}^{(0,2)}(r)\over r^2}{\bf L}_2^2 + V_r^{(0,2)}(r),
\ee
where ${\bf L}_1 \equiv {\bf r} \times {\bf p}_1$ and ${\bf L}_2 \equiv {\bf r} \times {\bf p}_2$. 
Note that neither ${\bf L}_1$ nor ${\bf L}_2$ corresponds to the orbital angular momentum 
of the particle and antiparticle. By using invariance under charge conjugation plus $m_1
\leftrightarrow m_2$ transformation, we obtain
\be
V_{{\bf p}^2}^{(2,0)}(r) =V_{{\bf p}^2}^{(0,2)}(r), \qquad 
V_{{\bf L}^2}^{(2,0)}(r) =V_{{\bf L}^2}^{(0,2)}(r), \qquad 
V_r^{(2,0)}(r)=V_r^{(0,2)}(r;m_2 \leftrightarrow m_1). 
\ee
The spin-dependent part of $V^{(2,0)}$ is of the type 
\be
V^{(2,0)}_{SD}=V^{(2,0)}_{LS}(r){\bf L}_1\cdot{\bf S}_1.
\ee
Analogously, for the $V^{(0,2)}$ potential we can write 
\be
V^{(0,2)}_{SD}=-V^{(0,2)}_{LS}(r){\bf L}_2\cdot{\bf S}_2. 
\ee
From invariance under charge conjugation plus $m_1 \leftrightarrow m_2$ transformation, we obtain  
$$
V^{(2,0)}_{LS}(r)=V^{(0,2)}_{LS}(r; m_2 \leftrightarrow m_1). 
$$
By using Eqs. (\ref{singlet}) and (\ref{En20}) we get, in terms of Wilson loop operators:  
\be
V_{{\bf p}^2}^{(2,0)}(r)={i \over 2}{\hat {\bf r}}^i{\hat {\bf r}}^j
\lim_{T\rightarrow \infty}\int_0^{T}dt \,t^2 \lla g{\bf E}_1^i(t) g{\bf E}_1^j(0) \rra_c,
\ee
\be
V_{{\bf L}^2}^{(2,0)}(r)={i \over 4}
\left(\delta^{ij}-3{\hat {\bf r}}^i{\hat {\bf r}}^j \right)
\lim_{T\rightarrow \infty}\int_0^{T}dt \, t^2 \lla g{\bf E}_1^i(t) g{\bf E}_1^j(0) \rra_c,
\ee
\bea
&&
V_r^{(2,0)}(r)= - {c_D^{(1)\prime} \over 8} 
\lim_{T_W \rightarrow \infty} \lla [{\bf D}_1,g{\bf E}_1](t) \rra_c 
\\
\nn
&&
- {i c_F^{(1)\,2} \over 4}  
\lim_{T\rightarrow  \infty}\int_0^{T}dt 
\lla g{\bf B}_1(t)\cdot g{\bf B}_1(0) \rra_c
+ {1 \over 2}(\bfnabla_r^2 V_{{\bf p}^2}^{(2,0)})
\\
\nn
&&
-{i \over 2}
\lim_{T\rightarrow \infty}\int_0^{T}dt_1\int_0^{t_1} dt_2 \int_0^{t_2}
dt_3\, (t_2-t_3)^2 \lla g{\bf E}_1(t_1)\cdot g{\bf E}_1(t_2) g{\bf E}_1(t_3)\cdot g{\bf E}_1(0) \rra_c 
\\
\nn
&&
+ {1 \over 2}
\left(\bfnabla_r^i
\lim_{T\rightarrow
  \infty}\int_0^{T}dt_1\int_0^{t_1} dt_2 \, (t_1-t_2)^2 \lla
g{\bf E}_1^i(t_1) g{\bf E}_1(t_2)\cdot g{\bf E}_1(0) \rra_c
\right)
\\
\nn
&&
- {i \over 2}
\left(\bfnabla_r^i V^{(0)}\right)
\lim_{T\rightarrow
  \infty}\int_0^{T}dt_1\int_0^{t_1} dt_2 \, (t_1-t_2)^3 \lla
g{\bf E}_1^i(t_1) g{\bf E}_1(t_2)\cdot g{\bf E}_1(0) \rra_c
\\
&&
\nn
- {1 \over 2}
\lim_{T\rightarrow \infty}\int_0^{T}dt_1\int_0^{t_1} dt_2 \, (t_1-t_2)^2 \lla
[{\bf D}_1.,g{\bf E}_1](t_1) g{\bf E}_1(t_2)\cdot g{\bf E}_1(0) \rra_c
\\
&&
\nn
+ {i \over 8}
\lim_{T\rightarrow \infty}\int_0^{T}dt \, t^2
\lla [{\bf D}_1.,g{\bf E}_1](t) [{\bf D}_1.,g{\bf E}_1](0) \rra_c
\\
&&
\nn
- {i \over 4}
\left(\bfnabla_r^i
\lim_{T\rightarrow \infty}\int_0^{T}dt \, t^2
\lla g{\bf E}_1^i(t) [{\bf D}_1.,g{\bf E}_1](0) \rra_c
\right)
\\
&&
\nn
- {1 \over 4}
\lim_{T\rightarrow \infty}\int_0^{T}dt \, t^3
\lla [{\bf D}_1.,g{\bf E}_1](t) g{\bf E}_1^j (0) \rra_c (\bfnabla_r^j
V^{(0)})
\\
&&
\nn
+ {1 \over 4}
\left(\bfnabla_r^i
\lim_{T\rightarrow \infty}\int_0^{T}dt \, t^3
\lla g{\bf E}_1^i(t) g{\bf E}_1^j (0) \rra_c (\bfnabla_r^j V^{(0)})
\right)
\\
&&
\nn
- {i \over 12}
\lim_{T\rightarrow \infty}\int_0^{T}dt \, t^4
\lla g{\bf E}_1^i(t) g{\bf E}_1^j (0) \rra_c
(\bfnabla_r^i V^{(0)}) (\bfnabla_r^j V^{(0)})
\\
& & 
- d_3^{(1)\prime} f_{abc} \int d^3{\bf x} \, \lim_{T_W \rightarrow \infty} 
g \lla G^a_{\mu\nu}({x}) G^b_{\mu\al}({x}) G^c_{\nu\al}({x}) \rra  \nn 
\eea
(note that, although, formally the first and last terms depend on the time where the 
operator insertion is made, this is not so after doing the $T_W \rightarrow
\infty$ limit\footnote{$V^{(0)}$ could also be written in a similar way:
$$
V^{(0)}={1\over 2}\int d^3{\bf x} \, \lim_{T_W \rightarrow \infty} 
\lla \left( {\bfPi}^a{\bfPi}^a +{\bf B}^a{\bf B}^a \right)({x}) \rra.
$$}),
\be
V_{LS}^{(2,0)}(r) = -{c_F^{(1)} \over r^2}i {\bf r}\cdot \lim_{T\rightarrow \infty}\int_0^{T}dt \, t \,  
\lla g{\bf B}_1(t) \times g{\bf E}_1 (0) \rra + {c_S^{(1)}\over 2 r^2}{\bf r}\cdot (\bfnabla_r V^{(0)}).
\label{vls20}
\ee

For the $V^{(1,1)}$ potential we define
\be
V^{(1,1)}=V^{(1,1)}_{SD}+V^{(1,1)}_{SI}.
\ee
The spin-independent part can be written as 
\be
V^{(1,1)}_{SI}= -{1 \over 2}\left\{{\bf p}_1\cdot {\bf p}_2,V_{{\bf p}^2}^{(1,1)}(r)\right\}
-{V_{{\bf L}^2}^{(1,1)}(r)\over 2r^2}({\bf L}_1\cdot{\bf L}_2+ {\bf L}_2\cdot{\bf L}_1)+ V_r^{(1,1)}(r),
\ee
while the spin-dependent part contains the following operators:
\be
V^{(1,1)}_{SD}=
V_{L_1S_2}^{(1,1)}(r){\bf L}_1\cdot{\bf S}_2 - V_{L_2S_1}^{(1,1)}(r){\bf L}_2\cdot{\bf S}_1
+ V_{S^2}^{(1,1)}(r){\bf S}_1\cdot{\bf S}_2 + V_{{\bf S}_{12}}^{(1,1)}(r){\bf S}_{12}({\hat {\bf r}}), 
\ee
where ${\bf S}_{12}({\hat {\bf r}}) \equiv 3 {\hat {\bf r}}\cdot \bfsigma_1 \,{\hat {\bf r}}\cdot \bfsigma_2 
- \bfsigma_1\cdot \bfsigma_2$.  
Because of the invariance under charge conjugation plus $m_1 \leftrightarrow m_2$ transformation, we have 
$$
V_{L_1S_2}^{(1,1)}(r)=V_{L_2S_1}^{(1,1)}(r; m_1 \leftrightarrow m_2). 
$$
By using Eqs. (\ref{singlet}) and (\ref{En11}) we get, in terms of Wilson loop operators:
\be
V_{{\bf p}^2}^{(1,1)}(r)=i{\hat {\bf r}}^i{\hat {\bf r}}^j
\lim_{T\rightarrow \infty}\int_0^{T}dt \, t^2
\lla g{\bf E}_1^i(t) g{\bf E}_2^j(0) \rra_c,
\ee
\be
V_{{\bf L}^2}^{(1,1)}(r)=i
{\delta^{ij}-3{\hat {\bf r}}^i{\hat {\bf r}}^j \over 2}
\lim_{T\rightarrow \infty}\int_0^{T}dt \,t^2
\lla g{\bf E}_1^i(t) g{\bf E}_2^j(0) \rra_c,
\ee
\bea
&&
V_r^{(1,1)}(r)= -{1 \over 2}(\bfnabla_r^2 V_{{\bf p}^2}^{(1,1)})
\\
\nn
&&
-i \lim_{T\rightarrow
  \infty}\int_0^{T}dt_1\int_0^{t_1} dt_2 \int_0^{t_2}
dt_3\, (t_2-t_3)^2 \lla g{\bf
  E}_1(t_1)\cdot g{\bf E}_1(t_2) g{\bf E}_2(t_3)\cdot g{\bf E}_2(0) \rra_c 
\\
\nn
&&
+
{1 \over 2}
\left(\bfnabla_r^i
\lim_{T\rightarrow
  \infty}\int_0^{T}dt_1\int_0^{t_1} dt_2 (t_1-t_2)^2 
\lla g{\bf E}_1^i(t_1) g{\bf E}_2(t_2)\cdot g{\bf E}_2(0) \rra_c \right)
\\
\nn
&&
+ {1 \over 2}
\left(\bfnabla_r^i
\lim_{T\rightarrow
  \infty}\int_0^{T}dt_1\int_0^{t_1} dt_2 (t_1-t_2)^2 
\lla g{\bf E}_2^i(t_1) g{\bf E}_1(t_2)\cdot g{\bf E}_1(0) \rra_c
\right)
\\
\nn
&&
- {i \over 2}
\left(\bfnabla_r^i V^{(0)}\right)
\lim_{T\rightarrow
  \infty}\int_0^{T}dt_1\int_0^{t_1} dt_2  (t_1-t_2)^3 
\lla g{\bf E}_1^i(t_1) g{\bf E}_2(t_2)\cdot g{\bf E}_2(0) \rra_c
\\
\nn
&&
- {i \over 2}
\left(\bfnabla_r^i V^{(0)}\right)
\lim_{T\rightarrow
  \infty}\int_0^{T}dt_1\int_0^{t_1} dt_2  (t_1-t_2)^3 
\lla g{\bf E}_2^i(t_1) g{\bf E}_1(t_2)\cdot g{\bf E}_1(0) \rra_c
\\
&&
\nn
- {1 \over 2}
\lim_{T\rightarrow
  \infty}\int_0^{T}dt_1\int_0^{t_1} dt_2  (t_1-t_2)^2 
\lla [{\bf D}_1.,g{\bf E}_1](t_1) g{\bf E}_2(t_2)\cdot g{\bf E}_2(0) \rra_c
\\
&&
\nn
+ {1 \over 2}
\lim_{T\rightarrow
  \infty}\int_0^{T}dt_1\int_0^{t_1} dt_2  (t_1-t_2)^2 
\lla [{\bf D}_2.,g{\bf E}_2](t_1) g{\bf E}_1(t_2)\cdot g{\bf E}_1(0) \rra_c
\\
&&
\nn
- {i \over 4}
\lim_{T\rightarrow \infty}\int_0^{T}dt \, t^2
\lla [{\bf D}_1.,g{\bf E}_1](t) [{\bf D}_2.,g{\bf E}_2](0) \rra_c
\\
&&
\nn
+ {i \over 4}
\left(\bfnabla_r^i
\lim_{T\rightarrow \infty}\int_0^{T}dt \, t^2
\left\{
\lla g{\bf E}_1^i(t) [{\bf D}_2.,g{\bf E}_2](0) \rra_c
-
\lla g{\bf E}_2^i(t) [{\bf D}_1.,g{\bf E}_1](0) \rra_c
\right\}
\right)
\\
&&
\nn
- {1 \over 4}
\lim_{T\rightarrow \infty}\int_0^{T}dt \, t^3
\left\{
\lla [{\bf D}_1.,g{\bf E}_1](t) g{\bf E}_2^j (0) \rra_c 
- \lla [{\bf D}_2.,g{\bf E}_2](t) g{\bf E}_1^j (0) \rra_c
\right\}
(\bfnabla_r^j
V^{(0)})
\\
&&
\nn
+ {1 \over 4}
\left(\bfnabla_r^i
\lim_{T\rightarrow \infty}\int_0^{T}dt \, t^3
\left\{
\lla g{\bf E}_1^i(t) g{\bf E}_2^j (0) \rra_c 
+ \lla g{\bf E}_2^i(t) g{\bf E}_1^j (0) \rra_c
\right\} 
(\bfnabla_r^j V^{(0)})
\right)
\\
&&
\nn
- {i \over 6}
\lim_{T\rightarrow \infty}\int_0^{T}dt \, t^4
\lla g{\bf E}_1^i(t) g{\bf E}_2^j (0) \rra_c
(\bfnabla_r^i V^{(0)}) (\bfnabla_r^j V^{(0)})
\\
& &
+ (d_{ss} + d_{vs} \lim_{T_W\rightarrow \infty} \lla T_1^a T_2^{a} \rra ) 
\,\delta^{(3)}({\bf x}_1-{\bf x}_2) \nn
\eea
(here and in the following formulas the two colour matrices in 
$\lla T_1^a T_2^{a} \rra$ are inserted in the Wilson loop at the same time: $-T_W/2 \le t \le T_W/2$; 
the $t$ dependence disappears in the $T_W\to\infty$ limit),
\be
V_{L_2S_1}^{(1,1)}(r)= - {c_F^{(1)} \over r^2}i {\bf r}\cdot \lim_{T\rightarrow \infty}\int_0^{T}dt \, t \, 
\lla g{\bf B}_1(t) \times g{\bf E}_2 (0) \rra 
\label{vls11}
\,,
\ee
\bea
V_{S^2}^{(1,1)}(r)&=& {2 c_F^{(1)} c_F^{(2)} \over 3}i \lim_{T\rightarrow \infty}\int_0^{T} dt \,  
\lla g{\bf B}_1(t) \cdot g{\bf B}_2 (0) \rra
\label{vs2}
\\
\nn
&&
- 4(d_{sv} 
+ d_{vv} \lim_{T_W\rightarrow \infty}\lla T_1^a T_2^{a} \rra ) \,
\delta^{(3)}({\bf x}_1-{\bf x}_2)
\,,
\eea
\be
V_{{\bf S}_{12}}^{(1,1)}(r)=
{c_F^{(1)} c_F^{(2)} \over 4}i {\hat {\bf r}}^i{\hat {\bf r}}^j
\lim_{T\rightarrow \infty}\int_0^{T} dt \, 
\left[
\lla g {\bf B}^i_1(t) g {\bf B}^j_2 (0) \rra  - {\delta^{ij}\over 3}\lla g{\bf B}_1(t)
\cdot g{\bf B}_2 (0) \rra
\right].
\label{vs12}
\ee

We now compare our results with previous ones. For the
spin-dependent potentials we find agreement with the Eichten--Feinberg 
results \cite{spin1} (once the NRQCD matching coefficients have been
taken into account) {\it except} for the $1/m_1m_2$ spin-orbit potential
$V_{L_2S_1}^{(1,1)}$.  Since the Eichten--Feinberg results have been
checked by, at least, three independent groups \cite{spin2,BMP,chen}, we
perform a more detailed comparison in Appendix \ref{comparison}. We show that
our expression in terms of Wilson loops and theirs give different results in
terms of intermediate states and, more important, we show that they give
different perturbative results at leading order in $\als$.  Ours coincides
with the well-known tree-level calculation, whereas the
Eichten--Feinberg expression gives $1/2$ the expected result.
Moreover, our perturbative result fulfils the Gromes relation \cite{spin2}.
The fact that the same mistake has been done by several groups can only be
explained by a systematic error. We believe that their systematic error has to
do with the common assumption in the literature that one may neglect, {\it in general}, 
the dependence of the Wilson loops on the gluonic strings, or on any
other gluonic operator, at $t=\pm T_W/2$. An analysis of the calculation done
by Eichten and Feinberg in \cite{spin1} supports this belief. 
Finally, we would like to mention that several different expressions for
the spin-dependent potentials, in particular the correct one, can be found
in the literature dealing with the lattice evaluation of them 
\cite{camp,mich2,zerwas,gunnar,latpot}. All these refer to the
work of Eichten and Feinberg \cite{spin1} for the derivation. We believe that
our result makes mandatory a clarification of all previous lattice
evaluations of the spin-dependent potentials.

The spin-independent potentials have only been computed before by Barchielli,
Brambilla, Montaldi and Prosperi in \cite{BMP} (the analysis done in
\cite{thesis}, which appears to be inconclusive, has never been published). We
agree (once the NRQCD matching coefficients have been taken into account) with
their results for the momentum-dependent terms, but not for the
momentum-independent terms, where we find new contributions.  Moreover, since
the potential we get here is {\it complete} up to order $1/m^2$, it is not
affected by the ordering ambiguity, which affects the derivation in
\cite{BMP}.  In this context, we would like to mention that our result may be
of particular relevance for the study of the properties of the QCD vacuum in
the presence of heavy sources.  So far the lattice data for the spin-dependent
and spin-independent potentials are consistent with a flux-tube picture,
whereas it is only for the spin-dependent terms that the so-called ``scalar confinement'' is
consistent with lattice data \cite{rev0,flux} (however the lattice data are 
still not conclusive). It will be interesting to see how 
these pictures compare with the new momentum- and spin-independent potentials, once
lattice data will be available for them. We note that some of them are not
simply expressed by two field insertions on a static Wilson loop, such as the spin-
or the momentum-dependent terms.  In particular, an extended object coming from
the Yang--Mills sector is required (similar extended objects would also show
up by taking into account operators with light quarks).

\subsection{Gauss law and further identities}
The above results may be simplified and rewritten in several ways. 
For instance, by using the quantum-mechanical identities a)--c) given in section \ref{secqm}, we obtain 
\be
\lim_{T_W\rightarrow \infty} \lla [{\bf D}_1,g{\bf E}_1](t) \rra_c 
= - \left(\bfnabla_r^2 V^{(0)} + 2i\lim_{T\rightarrow \infty}\int_0^{T}dt \,
\lla g{\bf E}_1(t)\cdot g{\bf E}_1(0) \rra_c \right),
\ee
changing the expression of the Darwin term (that now looks similar to the
analogous expression given in Ref. \cite{BMP}). In fact, by using the 
quantum-mechanical identities a)--c) of section \ref{secqm}, we could systematically transform  
$[{\bf D},g{\bf E}]$ in terms of normal derivatives acting on matrix elements or on static energies.  

Another possibility, which turns out to be more powerful, is the use of the Gauss law (\ref{gausslaw}). 
It allows us to write all the terms of the type $[{\bf D},g{\bf E}]$ in terms of
$\delta^{(3)}({\bf x}_1-{\bf x}_2)$ times some color matrices (up to some terms
proportional to $\delta^{(3)}(0)$ that vanish in dimensional regularization).
More information can be obtained by using the behaviour of the Wilson loops
(or of the states) at
short distances for the terms proportional to the deltas (assuming they are
regular enough). It follows that all the original terms
with $[{\bf D},g{\bf E}]$ disappear except the Darwin term. Moreover, we have
($C_f=(N_c^2-1)/(2N_c)$)
$$
\lim_{T_W\rightarrow \infty}\lla T_1^a T_2^{a}\rra \delta^{(3)}({\bf x}_1-{\bf x}_2)
=C_f\delta^{(3)}({\bf x}_1-{\bf x}_2).
$$
Therefore, some potentials get simplified into the following expressions
\bea
&&
V_r^{(2,0)}(r)=  {\pi C_f \als c_D^{(1)\prime} \over 2} 
\delta^{(3)}({\bf x}_1-{\bf x}_2)
\\
\nn
&&
- {i c_F^{(1)\,2} \over 4}  
\lim_{T\rightarrow  \infty}\int_0^{T}dt 
\lla g{\bf B}_1(t)\cdot g{\bf B}_1(0) \rra_c
+ {1 \over 2}(\bfnabla_r^2 V_{{\bf p}^2}^{(2,0)})
\\
\nn
&&
-{i \over 2}
\lim_{T\rightarrow
  \infty}\int_0^{T}dt_1\int_0^{t_1} dt_2 \int_0^{t_2}
dt_3\, (t_2-t_3)^2 \lla g{\bf
  E}_1(t_1)\cdot g{\bf E}_1(t_2) g{\bf E}_1(t_3)\cdot g{\bf E}_1(0) \rra_c 
\\
\nn
&&
+ {1 \over 2}
\left(\bfnabla_r^i
\lim_{T\rightarrow
  \infty}\int_0^{T}dt_1\int_0^{t_1} dt_2 \, (t_1-t_2)^2 \lla
g{\bf E}_1^i(t_1) g{\bf E}_1(t_2)\cdot g{\bf E}_1(0) \rra_c
\right)
\\
\nn
&&
- {i \over 2}
\left(\bfnabla_r^i V^{(0)}\right)
\lim_{T\rightarrow
  \infty}\int_0^{T}dt_1\int_0^{t_1} dt_2 \, (t_1-t_2)^3 \lla
g{\bf E}_1^i(t_1) g{\bf E}_1(t_2)\cdot g{\bf E}_1(0) \rra_c
\\
&&
\nn
+ {1 \over 4}
\left(\bfnabla_r^i
\lim_{T\rightarrow \infty}\int_0^{T}dt\, t^3
\lla g{\bf E}_1^i(t) g{\bf E}_1^j (0) \rra_c (\bfnabla_r^j V^{(0)})
\right)
\\
&&
\nn
- {i \over 12}
\lim_{T\rightarrow \infty}\int_0^{T}dt \,t^4
\lla g{\bf E}_1^i(t) g{\bf E}_1^j (0) \rra_c
(\bfnabla_r^i V^{(0)}) (\bfnabla_r^j V^{(0)})
\\
& & 
- d_3^{(1)\prime} f_{abc} \int d^3{\bf x} \, \lim_{T_W \rightarrow \infty} 
g \lla G^a_{\mu\nu}({x}) G^b_{\mu\al}({x}) G^c_{\nu\al}({x}) \rra  \nn
\,,
\eea
\bea
&&
V_r^{(1,1)}(r)=
-{1 \over 2}(\bfnabla_r^2 V_{{\bf p}^2}^{(1,1)})
\\
\nn
&&
-i \lim_{T\rightarrow
  \infty}\int_0^{T}dt_1\int_0^{t_1} dt_2 \int_0^{t_2}
dt_3\, (t_2-t_3)^2 \lla g{\bf
  E}_1(t_1)\cdot g{\bf E}_1(t_2) g{\bf E}_2(t_3)\cdot g{\bf E}_2(0) \rra_c 
\\
\nn
&&
+
{1 \over 2}
\left(\bfnabla_r^i
\lim_{T\rightarrow
  \infty}\int_0^{T}dt_1\int_0^{t_1} dt_2 (t_1-t_2)^2 
\lla g{\bf E}_1^i(t_1) g{\bf E}_2(t_2)\cdot g{\bf E}_2(0) \rra_c \right)
\\
\nn
&&
+ {1 \over 2}
\left(\bfnabla_r^i
\lim_{T\rightarrow
  \infty}\int_0^{T}dt_1\int_0^{t_1} dt_2 (t_1-t_2)^2 
\lla g{\bf E}_2^i(t_1) g{\bf E}_1(t_2)\cdot g{\bf E}_1(0) \rra_c
\right)
\\
\nn
&&
- {i \over 2}
\left(\bfnabla_r^i V^{(0)}\right)
\lim_{T\rightarrow
  \infty}\int_0^{T}dt_1\int_0^{t_1} dt_2  (t_1-t_2)^3 
\lla g{\bf E}_1^i(t_1) g{\bf E}_2(t_2)\cdot g{\bf E}_2(0) \rra_c
\\
\nn
&&
- {i \over 2}
\left(\bfnabla_r^i V^{(0)}\right)
\lim_{T\rightarrow
  \infty}\int_0^{T}dt_1\int_0^{t_1} dt_2  (t_1-t_2)^3 
\lla g{\bf E}_2^i(t_1) g{\bf E}_1(t_2)\cdot g{\bf E}_1(0) \rra_c
\\
&&
\nn
+ {1 \over 4}
\left(\bfnabla_r^i
\lim_{T\rightarrow \infty}\int_0^{T}dt \, t^3
\left\{
\lla g{\bf E}_1^i(t) g{\bf E}_2^j (0) \rra_c 
+ \lla g{\bf E}_2^i(t) g{\bf E}_1^j (0) \rra_c
\right\} 
(\bfnabla_r^j V^{(0)})
\right)
\\
&&
\nn
- {i \over 6}
\lim_{T\rightarrow \infty}\int_0^{T}dt \, t^4
\lla g{\bf E}_1^i(t) g{\bf E}_2^j (0) \rra_c
(\bfnabla_r^i V^{(0)}) (\bfnabla_r^j V^{(0)})
+ (d_{ss} + d_{vs} C_f) \,\delta^{(3)}({\bf x}_1-{\bf x}_2) \nn,
\eea
\be
V_{S^2}^{(1,1)}(r)= {2 c_F^{(1)} c_F^{(2)} \over 3}i \lim_{T\rightarrow \infty}\int_0^{T} dt \,  
\lla g{\bf B}_1(t) \cdot g{\bf B}_2 (0) \rra
- 4(d_{sv} + d_{vv} C_f ) \,
\delta^{(3)}({\bf x}_1-{\bf x}_2).
\label{vs2sim}
\ee
Similar considerations also apply to the results in terms of states of section \ref{secqm}.

\section{Power counting}
\label{secpc}
The standard power counting of NRQCD (organized in powers of $v$ and $\als$)
used to assess the relative importance of the different matrix elements, as
discussed, for instance, in \cite{pc}, can only be proved in the perturbative
regime. Even in this regime, owing to the different dynamical scales still
involved, the matrix elements of NRQCD do not have a unique power counting in
$v$. In the non-perturbative regime the problem of the power counting of NRQCD
is still open.  In principle, it is possible that a different power counting
may be appropriate in this situation and this would influence, for
instance, the studies of the charmonium system or of higher bottomonium
states\footnote{ A different, non-standard, power counting of the matrix
  elements of NRQCD may explain the apparent difficulties that NRQCD is facing
  to explain the polarization of prompt $J/\psi$ data, and to accurately
  determine the different matrix elements (see \cite{Kraemer}).}.  We believe
that our result, through the connection between NRQCD and the
quantum-mechanical picture, will eventually help to better understand the
hierarchy of the different matrix elements in NRQCD, as well as to get a much
deeper understanding of the underlying dynamics. This is due to the fact that,
by going to a NR quantum-mechanical formulation, we have made the
dynamics of the heavy quarks explicit transfering the problem of the power counting of
NRQCD into the problem of obtaining the power counting of the different
potentials in pNRQCD.  These may be expressed in terms of Wilson loops where
only gluons and light quarks appear as dynamical entities and for which there
are or there will be direct lattice measurements. Moreover, it is in this
formulation that statements such as the virial theorem have a more rigorous,
gauge-independent meaning.

Here, we only say a few words about the expected behaviour of the potentials
using arguments of naturalness on the scale $mv$, i.e. assuming that the
potentials scale with $mv$.  We first consider $V^{(0)}$. In principle,
$V^{(0)}$ counts as $mv$, but, by definition, the kinetic energy counts as
$mv^2$. Therefore, the virial theorem constrains $V^{(0)}$ also to count as
$mv^2$.  The extra $O(v)$ suppression has to come on dynamical grounds.  In
the perturbative case, it originates from the factor $\als \sim v$ in the
potential.  In the non-perturbative case little can be said and other
mechanisms must be responsible. Using naturalness, $V^{(1,0)}/m$ scales like $mv^2$. Therefore, it could in principle be as
large as $V^{(0)}$. This makes a lattice calculation or a model evaluation of this potential urgent.
Perturbatively, owing to the factor $\als^2$, it is of $O(mv^4)$.
For what concerns the $1/m^2$ potentials, the naturalness argument suggests that they are of $O(mv^3)$. 
However, also here several constraints apply. Terms involving $\bfnabla
V^{(0)} \sim m^2v^3$ are suppressed by an extra factor $v$, due to the virial
theorem. The Gromes relation \cite{spin2,ChenKuang}, 
\be 
{1\over 2r}{d V^{(0)}\over dr}  +  V_{LS}^{(2,0)} - V_{L_2S_1}^{(1,1)} = 0,
\label{gromrel}
\ee 
suppresses by an extra factor $v$ the combination $V_{LS}^{(2,0)}$ $-$ $V_{L_2S_1}^{(1,1)}$. 
Similar constraints also exist for the spin-independent potentials \cite{BMP}.
Perturbatively the $1/m^2$ potentials count at most as $O(mv^4)$, because of the extra 
$\als$ suppression.  Finally, it is important to consider that some of the potentials 
are $O(\als)$-suppressed because of the matching coefficients inherited from
NRQCD. This is, for instance, 
the case of the terms coming from the $1/m^2$ corrections to the purely gluonic sector 
of the NRQCD Lagrangian or of the terms coming from the 4-fermion sector.  

Terms involving two field-strength insertions in the static Wilson loop 
are known from lattice measurements \cite{latpot} and have been studied in some 
QCD vacuum models \cite{mod}. For them a parameterization is possible and some 
supplementary information can be extracted. However, terms involving more than two field insertions 
in the static Wilson loop have not been studied so far, to our knowledge, by lattice 
simulations or within models. Consistency with the experimental data 
will further constrain any possible power-counting rule. 
In any case, a detailed study of the potentials using
the above information (as well as new lattice or model-dependent
results) should be performed in order to obtain the size (and thus the
power-counting rules) of the different potentials for the charmonium and
bottomonium systems.

\section{Conclusions and outlook}
\label{conclusions}
A new formalism with which to obtain the QCD potential at arbitrary orders in $1/m$
has been explained in detail. We have obtained expressions for the energies of the 
gluonic excitations between heavy quarks valid beyond perturbation theory at $O(1/m^2)$.
In particular, for the heavy quarkonium, we have also obtained the complete
spin-dependent and spin-independent potentials at $O(1/m^2)$ for pure
gluodynamics in terms of Wilson loops. For the spin-dependent piece our
results correct the expressions given in \cite{spin1,spin2,BMP}. For the
spin-independent potentials, we agree with the momentum-dependent potentials obtained 
in \cite{BMP}, but not for the momentum-independent terms, where new
contributions are found. We have also briefly discussed the power counting in the non-perturbative regime. 

We conclude, commenting on two possible developments of the present work.
First, it is worthwhile to explore the possibility of expressing the potentials 
associated with higher gluonic excitations in terms of Wilson loop operators as done here 
for the heavy quarkonium ground state. The corresponding quantum-mechanical expressions  
are given in Eqs. (\ref{Em1})--(\ref{En11}). 
Second, our results are complete at $O(1/m^2)$ in the case of pure gluodynamics. If we
want to incorporate light fermions, the procedure to be followed is analogous and our results still remain 
valid (considering now matrix elements and Wilson loops with dynamical light fermions incorporated), 
except for new terms appearing in the energies at $O(1/m^2)$ due to operators
involving light fermions that appear in the NRQCD Lagrangian at $O(1/m^2)$ \cite{ManBauer}. 
They may be incorporated along the same lines as the terms discussed here and will 
be explicitly worked out elsewhere.

\bigskip

{\bf Acknowledgements.} 
We thank Joan Soto for discussions.  
A.V. acknowledges Nora Brambilla for several discussions and comments 
and a discussion with Dieter Gromes on the spin-dependent potential issue. 
A.P. acknowledges the TMR contract No. ERBFMBICT983405.
A.V. thanks the CERN Theory Division for hospitality during the first 
stage of this work and the Alexander von Humboldt Foundation for 
support.

\vfill\eject

\appendix
\section{The parameterization of \protect\cite{BMP}}
In this appendix, for ease of comparison, we write the spin-independent potentials
in the parameterization given in \cite{BMP}. They read 
\be
V^{(2,0)}_{SI}={1 \over 2}\left\{{\bf p}_1^i{\bf p}_1^j,\delta^{ij}V_d(r)+\left({\delta^{ij} \over 3} 
-{\hat {\bf r}}^i{\hat {\bf r}}^j\right)V_e(r)\right\}
+ {1 \over 8}(\bfnabla_r^2[V^{(0)}+V_a]),
\ee
\be
V^{(1,1)}_{SI}={1 \over 2}\left\{{\bf p}_1^i{\bf p}_2^j,\delta^{ij}V_b(r)+\left({\delta^{ij} \over 3} 
-{\hat {\bf r}}^i{\hat {\bf r}}^j\right)V_c(r)\right\}
+ V_f(r) \,.
\ee
Let us note that in \cite{BMP}
the $1/m_1m_2$ potential contained only momentum-dependent pieces. 
Therefore, the momentum-independent potential, which we name $V_f$, was missing. 
Our calculation also substantially modifies the result for $V_a$ given in \cite{BMP}.
The above potentials read
\be
V_d={i \over 6}\lim_{T\rightarrow \infty}\int_0^{T}dt \, t^2
\lla g{\bf E}_1(t) \cdot g{\bf E}_1(0) \rra_c,
\ee
\be
\left({\delta^{ij} \over 3} -{\hat {\bf
        r}}^i{\hat {\bf r}}^j\right)V_e={i \over 2}\lim_{T\rightarrow
    \infty}\int_0^{T}dt \, t^2
\left\{ \lla g{\bf E}_1^i(t) g{\bf E}_1^j(0) \rra_c
- {\delta^{ij} \over 3} \lla g{\bf E}_1(t) \cdot g{\bf E}_1(0) \rra_c \right\},
\ee
\be
{1 \over 8}(\bfnabla_r^2[V^{(0)}+V_a]) = V_r^{(2,0)}(r)-
{1 \over 2}(\bfnabla_r^2 V_{{\bf p}^2}^{(2,0)}) 
+ {i\over 4} \left( \nabla_r^i\nabla_r^j \lim_{T\rightarrow
    \infty}\int_0^{T}dt \, t^2 \lla g{\bf E}_1^i(t) g{\bf E}_1^j(0) \rra_c \right),
\ee
\be
V_b=-{i \over 3}\lim_{T\rightarrow \infty}\int_0^{T}dt \,t^2
\lla g{\bf E}_1(t) \cdot g{\bf E}_2(0) \rra_c,
\ee
\be
\left({\delta^{ij} \over 3} -{\hat {\bf
        r}}^i{\hat {\bf r}}^j\right)V_c=-i\lim_{T\rightarrow
    \infty}\int_0^{T}dt \,t^2
\left\{ \lla g{\bf E}_1^i(t) g{\bf E}_2^j(0) \rra_c
-
{\delta^{ij} \over 3} \lla g{\bf E}_1(t) \cdot g{\bf E}_2(0) \rra_c
\right\},
\ee
\be
V_f(r)= V_r^{(1,1)}(r)+
{1 \over 2}(\bfnabla_r^2 V_{{\bf p}^2}^{(1,1)})
+ {i\over 2} \left( \nabla_r^i\nabla_r^j \lim_{T\rightarrow
    \infty}\int_0^{T}dt \, t^2 \lla g{\bf E}_1^i(t) g{\bf E}_2^j(0) \rra_c \right).
\ee

\section{Comparison with the Eichten--Feinberg spin-orbit potential}
\label{comparison}
In order to compare our results with the Eichten--Feinberg ones properly, we set $c_F^{(1)}=c_F^{(2)}=1$. 
Then, our Eq. (\ref{vls11}) reads
\bea
\label{nos}
V_{L_2S_1}^{(1,1)}(r)&=& 
-{i \over r^2} {\bf r}\cdot \lim_{T\rightarrow \infty}\int_0^{T}dt \, t \, 
\lla g{\bf B}_1(t) \times g{\bf E}_2 (0) \rra 
\\
\nn
&=&
{i \over r^2} {\bf r}\cdot
\sum_{k\neq 0} 
{^{(0)}\langle 0| g {\bf B}_1|k\rangle^{(0)}  \times\,
^{(0)}\langle k|g {\bf E}_2^{T}|0\rangle^{(0)} \over (E_0^{(0)} -
E_k^{(0)})^2}  
\\
\nn
&{\buildrel{=} \over {\hbox{\tiny pert}}}&
{C_f\als \over r^3} + O(\als^2).
\eea
On the other hand, Eichten and Feinberg obtain (we actually use the expression in Minkowski space 
given in Ref. \cite{BMP}):
\bea
\label{EF}
V_{L_2S_1}^{(1,1)}(r)
&=& 
{i \over 2 r^2} \lim_{T_W\rightarrow \infty}{1 \over T_W} 
\int_{-T_W/2}^{T_W/2}dt \, \int_{-T_W/2}^{T_W/2}dt' \,t'
{\bf r}\cdot  
\lla g{\bf B}_1(t) \times g{\bf E}_2 (t') \rra 
\\
\nn
&=&
{i \over 2r^2} {\bf r}\cdot
\sum_{m \neq 0} \sum_{k\neq 0,m} 
{a_0a_m \over a_0^2}
{
^{(0)}\langle 0| g {\bf B}_1|k\rangle^{(0)}  \times\,
^{(0)}\langle k|g {\bf E}_2^{T}|m\rangle^{(0)} 
\over (E_0^{(0)} - E_k^{(0)})(E_0^{(0)} - E_m^{(0)})
}
\\
\nn
&&
-{i \over 2r^2} {\bf r}\cdot
\sum_{m \neq 0} \sum_{k\neq 0,m} 
{a_0a_m \over a_0^2}
{
^{(0)}\langle m| g {\bf B}_1|k\rangle^{(0)}  \times\,
^{(0)}\langle k|g {\bf E}_2^{T}|0\rangle^{(0)} 
\over (E_0^{(0)} - E_k^{(0)})(E_0^{(0)} - E_m^{(0)})}
\\
\nn
&&
+{i \over 2r^2} {\bf r}\cdot
\sum_{k\neq 0} 
{^{(0)}\langle 0| g {\bf B}_1|k\rangle^{(0)}  \times\,
^{(0)}\langle k|g {\bf E}_2^{T}|0\rangle^{(0)} \over (E_0^{(0)} -
E_k^{(0)})^2}
\\
\nn
&{\buildrel{=} \over {\hbox{\tiny pert}}}&
{C_f\als \over 2r^3} + O(\als^2),
\eea
where the $a_n({\bf x}_1,{\bf x}_2)$ are defined by 
$$
\psi^\dagger({\bf x}_1)\phi({\bf x}_1,{\bf x}_2) \chi({\bf x}_2) \vert \vac \rangle = 
\sum_n a_n({\bf x}_1,{\bf x}_2)\vert \underbar{n};{\bf x}_1,{\bf x}_2 \rangle^{(0)},  
$$
being
\begin{equation}
\phi({\bf y},{\bf x}) \equiv {\rm P} \exp \left\{ ig \displaystyle 
\int_0^1 \!\! ds \, ({\bf y} - {\bf x}) \cdot {\bf A}({\bf x} - s({\bf x} - {\bf y})) \right\}
\label{schwinger}
\end{equation}
the end-point string used in the Wilson loop operators. Note that we have
fixed $T_W=T$ in Eq. (\ref{EF}), as corresponds to the procedure followed by
Eichten and Feinberg. 

The above calculation makes manifest the disagreement of Eq. (\ref{nos}) with Eq. (\ref{EF}) 
both at the perturbative level as well as in the representation in terms of intermediate states.  
A possible source of disagreement may be traced back in the original paper of Eichten and
Feinberg \cite{spin1} to their Eq. (4.9b), which seems to be incorrect.
Finally, the reason of this last error seems to be the improper treatment of
the Wilson loops in the large-time limit.


\begin{thebibliography}{999}
\bibitem{rev0} N. Brambilla and A. Vairo, hep-ph/9904330.
\bibitem{rev1} F. J. Yndur\'ain, hep-ph/9910399.
\bibitem{gunnar} G. S. Bali,  hep-ph/0001312.   
\bibitem{pNRQCD} A. Pineda and J. Soto, Nucl. Phys. {\bf B} (Proc. Suppl.) {\bf 64}, 428 (1998); 
 N. Brambilla, A. Pineda, J. Soto and A. Vairo, Phys. Rev. {\bf D 60}, 091502 (1999); 
 Nucl. Phys. {\bf B 566}, 275 (2000).  
\bibitem{m1} N. Brambilla, A. Pineda, J. Soto and A. Vairo, Phys. Rev. {\bf D 63}, 
 014023 (2001). 
\bibitem{wilson} K. G. Wilson, Phys. Rev. {\bf D 10}, 2445 (1974). 
\bibitem{Brown} L. Susskind, in {\it Les Houches, Session XXIX}, 
 ed. R. Balian and C. H. Llewellyn Smith (North-Holland Publishing Company, 
 Amsterdam, 1977); W. Fischler, Nucl. Phys. {\bf B 129}, 157 (1977);
 L. S. Brown and W.I. Weisberger, Phys. Rev. {\bf D 20}, 3239 (1979). 
\bibitem{spin1}  E. Eichten and F. L. Feinberg, Phys. Rev. {\bf D 23}, 2724
  (1981). 
\bibitem{Peskin} M. E. Peskin, in Proceeding of the 11th SLAC Institute, SLAC
  Report No. 207, 151, edited by P. Mc Donough (1983).  
\bibitem{spin2}  D. Gromes, Z. Phys. {\bf C 26}, 401 (1984).
\bibitem{thesis} R. Tafelmayer, Diploma Thesis, Heidelberg (1986). 
\bibitem{BMP}   A. Barchielli, E. Montaldi and G. M. Prosperi, Nucl. Phys. {\bf B 296}, 625 (1988); 
 (E) {\it ibid.} {\bf 303}, 752 (1988); 
 A. Barchielli, N. Brambilla and G. Prosperi, Nuovo Cimento {\bf 103 A}, 59 (1990). 
\bibitem{chen} Y. Chen, Y. Kuang and R. J. Oakes, Phys. Rev. {\bf D 52}, 264 (1995).
\bibitem{latpot} G. S. Bali, K. Schilling and A. Wachter, Phys. Rev. {\bf D 55}, 5309 (1997); 
 {\bf D 56}, 2566 (1997). 
\bibitem{BV1} N. Brambilla and A. Vairo, Nucl. Phys. {\bf B} (Proc. Suppl.) {\bf 74}, 210 (1999). 
\bibitem{mod} N. Brambilla, P. Consoli and G. M. Prosperi, Phys. Rev. {\bf D 50}, 5878 (1994); 
 N. Brambilla and A. Vairo, Phys. Rev. {\bf D 55}, 3974 (1997); 
 M. Baker, J. S. Ball, N. Brambilla and A. Vairo, Phys. Lett. {\bf B 389}, 577 (1996). 
\bibitem{SS} A. P. Szczepaniak and E. S. Swanson, Phys. Rev. {\bf D 55}, 3987 (1997).
\bibitem{NRQCD} W. E. Caswell and G. P. Lepage, Phys. Lett. {\bf B 167}, 437 (1986); 
 G. T. Bodwin, E. Braaten and G. P. Lepage, Phys. Rev. {\bf D 51}, 1125 (1995);  
 Erratum, {\it ibid.} {\bf D 55}, 5853 (1997).
\bibitem{Manohar} A. V. Manohar, Phys. Rev. {\bf D 56}, 230  (1997).
\bibitem{Match} A. Pineda and J. Soto, Phys. Rev. {\bf D 58}, 114011 (1998).
\bibitem{others} P. Labelle, Phys. Rev. {\bf D 58}, 093013 (1998); 
 M. Beneke and V. A. Smirnov, Nucl. Phys. {\bf B 522}, 321 (1998); 
 H. W. Griesshammer, Phys. Rev. {\bf D 58}, 094027 (1998); 
 M. E. Luke, A. V. Manohar and I. Z. Rothstein, Phys. Rev. {\bf D 61}, 074025
 (2000). 
\bibitem{bcexp} F. Abe et al., CDF Collaboration, Phys. Rev. Lett. {\bf 81}, 2432 (1998).
\bibitem{bc} N. Brambilla and A. Vairo, hep-ph/0002075 and references therein.
\bibitem{ManBauer} C. Bauer and A. V. Manohar, Phys. Rev. {\bf D 57}, 337 (1998). 
\bibitem{CMYm6} A. Czarnecki, K. Melnikov and A. Yelkhovsky, Phys. Rev. {\bf A 59}, 4316 (1999). 
\bibitem{ai1}  A. V. Manohar and I. W. Stewart, Phys. Rev. {\bf D 62}, 
 074015 (2000).
\bibitem{knetter} C. Grosse-Knetter, Phys. Rev. {\bf D 49}, 1988 (1994); {\bf D 49}, 6709 (1994).
\bibitem{michael}  C. Michael, hep-ph/9809211; K. J. Juge, J. Kuti and C. Morningstar, hep-lat/9809015. 
\bibitem{DS} D. Eiras and J. Soto, Phys. Rev. {\bf D 61}, 114027 (2000).
\bibitem{camp} M. Campostrini, Nucl. Phys. {\bf B 256}, 717 (1985).
\bibitem{mich2} C. Michael, Phys. Rev. Lett. {\bf 56}, 1219 (1986); A. Huntley
  and C. Michael, Nucl. Phys. {\bf B 286}, 211 (1987). 
\bibitem{zerwas} K. D. Born, E. Laermann, T. F. Walsh and P. M. Zerwas, Phys. Lett {\bf B 329}, 332 (1994). 
\bibitem{flux} N. Brambilla, hep-ph/9809263.
\bibitem{pc} G. P. Lepage, L. Magnea, C. Nakhleh, U. Magnea and K. Hornbostel, Phys. Rev. {\bf D 46}, 4052 (1992).
\bibitem{Kraemer} M. Kraemer, BEACH00 proceedings, hep-ph/0010137; 
 A.K. Leibovich, BEACH00
  proceedings, hep-ph/0008236.
\bibitem{ChenKuang} Y. Chen and Y. Kuang, Z. Phys. {\bf C 67}, 627 (1995).
\end{thebibliography}
\end{document}